\documentclass[usegraphicx]{mn2e}
\usepackage{subfigure}
\newcommand{\rsun}{$\mathrm{R_{\sun}}$}
\newcommand{\msun}{$\mathrm{M_{\sun}}$}

\newcommand{\mdot}{$\dot{M}$}

\newcommand{\rate}{$\mathrm{M_{\sun}} \, \mathrm{yr}^{-1}$}

\newcommand{\xpn}[2]{$#1 \times 10^{#2}$}
\newcommand{\kms}{$\mathrm{km \, s^{-1}}$}

\title[The late stages of helium star-neutron star binaries]
      {The late stages of evolution of helium star-neutron star binaries 
       and the formation of double neutron star systems}
\author[Dewi \& Pols]
       {J. D. M. Dewi$^{1,3}$\thanks{email: jasinta@astro.uva.nl (JDMD), 
                                            O.R.Pols@astro.uu.nl (ORP)}, 
        O. R. Pols$^2$\footnotemark[1]\\
        $^1$Astronomical Institute {\it Anton Pannekoek},
            University of Amsterdam, Kruislaan 403, NL-1098 SJ Amsterdam,
            The Netherlands\\
        $^2$Astronomical Institute, University of Utrecht, 
            Postbus 80000, 3508 TA Utrecht, The Netherlands\\
        $^3$Bosscha Observatory and Department of Astronomy, Institut Teknologi Bandung, 
            Jl. Ganesha 10, Bandung 40132, Indonesia}
\date{Accepted . Received ; in original form }

\pagerange{\pageref{firstpage}--\pageref{lastpage}}
\pubyear{}

\begin{document}

\maketitle

\label{firstpage}

\begin{abstract}
With a view to understanding the formation of double neutron-star binaries, we investigate the late stages of evolution of helium stars with masses of 2.8 -- 6.4~\msun\ in binary systems with a 1.4~\msun\ neutron-star companion. We found that mass transfer from 2.8 -- 3.3~\msun\ helium stars (originating from main-sequence stars with masses of 10 -- 12~\msun\ which underwent case B evolution, or 9 -- 10~\msun\ which experienced case C mass transfer) as well as from 3.3 -- 3.8~\msun\ in very close orbits ($P_{\mathrm{orb}} \la 0\fd25$) will end up in a common-envelope and spiral-in phase due to the development of a convective helium envelope at the end of the calculation. If the neutron star has sufficient time to complete the spiraling-in process in the envelope of the helium star before the core collapses, the system will produce very tight double neutron-star binaries ($P_{\mathrm{orb}} \sim 0\fd01$) with a very short merger timescale, i.e. of the order of 1 Myr or less. These systems would have important consequences for the detection rate of gravitational-wave radiation and for the understanding of $\gamma$-ray burst progenitors. On the other hand, if the time left until the explosion is shorter than the orbital-decay timescale, the system will undergo a supernova explosion during the common-envelope phase.

Helium stars with masses 3.3 -- 3.8~\msun\ in wider orbits ($P_{\mathrm{orb}} \ga 0\fd25$) and those more massive than 3.8~\msun\ do not develop a convective envelope and therefore are not expected to go through common-envelope evolution. The remnants of these massive helium stars are double neutron-star pulsars with periods in the range of $0\fd1 - 1^{\mathrm{d}}$. This suggests that this range of mass (originating from main-sequence stars more massive than 12~\msun\ which underwent case B evolution, or more massive than 10~\msun\ which experienced case C mass transfer) includes the progenitors of the galactic double neutron-star pulsars with close orbits (B1913+16 and B1534+12). A minimum kick velocity of 70~\kms\ and 0~\kms\ (for B1913+16 and B1534+12, respectively) must have been imparted at the birth of the pulsar's companion. The double neutron stars with wider orbits (J1518+4904 and probably J1811-1736) are produced from helium star-neutron star binaries which avoid Roche-lobe overflow, with the helium star more massive than 2.5~\msun, i.e. the remnants of main-sequence stars more massive than 10~\msun\ in relatively wide orbits. For these systems the minimum kick velocities are 50~\kms\ and 10~\kms\ (for J1518+4904 and J1811-1736, respectively).
\end{abstract}

\begin{keywords}
stars: evolution -- binaries: general -- stars: neutron  -- 
pulsars: individual: B1913+16, B1534+12, J1518+4904, J1811-1736 
-- supernovae: general
\end{keywords}


\section{Introduction}
\label{doublens:sec:intro}

In a work published earlier (Dewi et al. 2002, hereafter called Paper~I), we discussed two types of mass-transfer evolution from a helium star to a neutron star, i.e. case BA (in which the onset of mass transfer occurs during helium core burning) and case BB (the star fills its Roche lobe after helium core burning is terminated, but before the ignition of carbon). We found that dynamically stable case BA mass transfer can take place from helium stars less massive than 3.0~\msun. The remnants of this type of evolution are heavy CO white dwarfs. Case BB mass transfer from helium stars less massive than 2.6 -- 2.8~\msun\ produces white dwarfs (CO white dwarf for masses $\la 1.9 - 2.0$~\msun/ helium stars or ONeMg white dwarf for larger masses), while more massive helium stars leave neutron stars as their remnants\footnote{These limiting masses depend on the initial orbital period of the system.}. In Paper~I we suggested that the systems with relatively high-mass helium stars and/or wide orbits are progenitors of type Ib supernova, and lower-mass helium stars or systems in close orbits produce type Ic supernovae. We were also able to constrain the possible progenitors of the observed galactic double neutron star (DNS) pulsars B1913+16 and B1534+12, if we assume that these DNSs were produced from helium star-neutron star binaries which went through a mass-transfer phase.

The calculations in Paper~I were done up to various stages of evolution, i.e. ranging from the appearance of the first carbon-burning convective shell to the outward penetration of the helium burning convective shell into the helium envelope. Although we could not follow the evolution further, we are interested to investigate the possible outcomes that the systems might produce, e.g. whether a common envelope (CE) and spiral-in will occur, whether the system will survive the CE phase, whether the supernova (SN) explosion will take place after or before the neutron star completes the spiraling-in process, and which type of SN the system will produce. We also want to find the possible progenitors of the wide-orbit DNS J1518+4904 and to put more constraints on the progenitors of B1913+16 and B1534+12. These are the problems we will discuss in this paper.

As the completion to Paper~I we first evolved helium stars in wider binary systems with a 1.4~\msun\ neutron-star companion, such that Roche-lobe overflow (RLOF) is initiated during carbon core burning or beyond (which we call case BC mass transfer). A study of this particular type of evolution has been done by Habets (1986) who evolved a 2.5~\msun\ helium star with a 17~\msun\ main-sequence companion in a period of $20\fd25$. Because we are mainly interested to study the systems which will produce DNS, we limit our calculations to helium stars in the mass range of 2.8 -- 6.4~\msun. Helium stars more massive than 6.4~\msun\ do not expand very much, so that their evolution in wide orbits will be similar to the evolution of a single helium star.

We discuss the results of case BC evolution in Sect.~\ref{doublens:sec:results}; the possible remnants of case BB and case BC evolution in Sect.~\ref{doublens:sec:remnants}; and the formation of the observed galactic DNSs in Sect.~\ref{doublens:sec:doublens}. The conclusions are given in Sect.~\ref{doublens:sec:conclusions}. Throughout the paper, in order to avoid confusion about the various stages of evolution, a subscript {\it i} is used to indicate the initial parameters (at the start of calculation), {\it o} the parameters at the end of the mass-transfer phase (at the initiation of the CE phase), {\it t} those at the end of the spiral-in phase (before SN explosion) and {\it f} the post-SN parameters.


\section{Results: Case BC Mass Transfer}
\label{doublens:sec:results}

\subsection{A brief description on the method of calculation}
\label{doublens:subsec:method}

We used the Eggleton code (Pols et al. 1995 and references therein) for the evolution of helium stars, assuming a chemical composition of (${Y} = 0.98, {Z} = 0.02$) and without enhanced mixing (the STD model in Pols (in preparation)). 

The non-conservative orbital evolution of the system is assumed to be affected by gravitational-wave radiation (Landau \& Lifshitz 1958) and by the loss of mass with angular momentum from the system (van den Heuvel 1994; Soberman, Phinney \& van den Heuvel 1997), such that
	\begin{eqnarray}
          \frac{\dot{a}}{a} & = & - \frac{64 \, G^{3}}{5 \, c^{5}} 
          \frac{M_{\mathrm{He}} \, M_{\mathrm{NS}} \, M_{\mathrm{T}}}{a^{4}} + 
          \frac{2 \, [\beta \, q^{2} - q + (\alpha - 1)]}{1 + q} 
          \frac{\dot{M}_{\mathrm{He}}}{M_{\mathrm{He}}} \nonumber \\ 
                            &   & - 2 \frac{\dot{M}_{\mathrm{NS}}}{M_{\mathrm{NS}}} + 
          \frac{\dot{M}_{\mathrm{T}}}{M_{\mathrm{T}}}
	  \label{doublens:eq:orbit}
	\end{eqnarray}
where $M_{\mathrm{He}}$, $M_{\mathrm{NS}}$, and $M_{\mathrm{T}}$ are the masses of the helium star and the neutron star, and the total mass of the system, respectively. $q = M_{\mathrm{He}} / M_{\mathrm{NS}}$ is the mass ratio of the system, $a$ is the orbital separation, $G$ is the constant of gravity, and $c$ the speed of light in vacuum. $\alpha$ is the fraction of mass lost from the helium star in the form of fast isotropic wind, and $\beta$ is the fraction of mass ejected isotropically from the vicinity of the neutron star. The stellar wind mass loss is given by eq.~(2) in Wellstein \& Langer (1999) multiplied by a factor of 0.5, i.e.
	\begin{eqnarray}
        \dot{M}_{\mathrm{He,wind}} & = & 
        \left\{ \begin{array}{ll} 
                2.8 \, 10^{-13} (L/\mathrm{L_{\sun}})^{1.5}, & 
                \log{L/\mathrm{L_{\sun}}} \geq 4.5 \\
                4.0 \, 10^{-37} (L/\mathrm{L_{\sun}})^{6.8}, & 
                \log{L/\mathrm{L_{\sun}}} < 4.5 
               \end{array} \right.
	\label{doublens:eq:wind}	 
	\end{eqnarray}
in \rate. We assumed that during the detached phase, the neutron star does not accrete matter from the stellar wind. During the mass transfer phase, the transferred matter is accreted up to the Eddington limit for helium accretion (\xpn{3}{-8}~\rate); the rest is lost from the system with the specific orbital angular momentum of the neutron star.

\setlength{\tabcolsep}{2.0pt}
	\begin{table*}  
         \caption[]{The binary parameters in case BC mass transfer from 2.8 -- 
                    6.4~\msun\ helium stars. The columns give: the initial mass and 
                    period; the duration, amount of mass removed from the helium star, 
                    and the maximum mass-loss rate during the Roche-lobe overflow; 
                    the mass and period at the end of the calculations, the final mass 
                    of helium in the 
                    envelope; and the final CO and ONeMg core masses. Masses are in 
                    solar units, the mass-loss rate in \rate, time in yr, and periods 
                    in days. The number in the {\it Note} column indicates during which 
                    consecutive convective carbon shell-burning phase the calculation 
                    stops: {\it a} implies the appearance of the shell, {\it b} during 
                    the shell burning, and {\it d} the disappearance of the shell.}
	 \label{doublens:tab:outcome}
	 \begin{center}
	 \begin{tabular}{rr|rrr|rrrccc}
	  \hline
	  \hline
          \noalign{\smallskip}
          $M_{\mathrm{i}}$ & $P_{\mathrm{i}}$~ & $\Delta t_{\mathrm{RLOF}}$~~ &
           $\Delta M$ & $\dot{M}_{\mathrm{max}}$~~~ & $M_{\mathrm{o}}$~ & 
          $P_{\mathrm{o}}$~ & $M_{\mathrm{He,e}}$ & $M_{\mathrm{CO}}$ & $M_{\mathrm{ONeMg}}$ 
          & Note \\
          \noalign{\smallskip}
	  \hline
          \noalign{\smallskip}
          2.8 &  2.0 & \xpn{6.16}{3} & 0.971 & \xpn{2.4}{-4} & 1.749 &  1.82 & 0.1481 & 
               1.438 & 1.348 & 4a \\
              &  5.0 & \xpn{3.42}{3} & 0.342 & \xpn{3.0}{-4} & 2.372 &  4.66 & 0.7537 & 
               1.433 & 1.360 & 4d \\
              & 10.0 & \xpn{4.74}{2} & 0.023 & \xpn{8.8}{-5} & 2.684 & 10.37 & 1.0578 & 
               1.436 & 1.378 & 5a \\
          2.9 &  2.0 & \xpn{4.58}{3} & 0.774 & \xpn{3.4}{-4} & 2.034 &  1.72 & 0.3576 & 
               1.470 & 1.371 & 4a \\
              &  4.0 & \xpn{2.41}{3} & 0.149 & \xpn{1.6}{-4} & 2.656 &  3.92 & 0.9676 & 
               1.478 & 1.401 & 4a \\
              &  6.0 & \xpn{2.57}{2} & 0.008 & \xpn{4.7}{-5} & 2.791 &  6.27 & 1.1039 & 
               1.498 & 1.313 & 3b \\
          3.2 &  1.0 & \xpn{3.55}{3} & 0.564 & \xpn{3.2}{-4} & 2.519 &  0.82 & 0.6418 & 
               1.640 & 1.446 & 3d \\
              &  2.0 & \xpn{1.22}{3} & 0.056 & \xpn{8.2}{-5} & 3.020 &  2.05 & 1.1305 & 
               1.626 & 1.432 & 3d \\
          \noalign{\smallskip}
	  \hline
          \noalign{\smallskip}
          3.6 &  0.6 & \xpn{3.32}{3} & 0.554 & \xpn{3.7}{-4} & 2.835 &  0.48 & 0.7365 & 
               1.961 & 1.597 & 3d \\
              &  1.2 & \xpn{8.67}{2} & 0.036 & \xpn{7.6}{-5} & 3.344 &  1.28 & 1.2380 & 
               1.963 & 1.394 & 3a \\
          4.0 &  0.5 & \xpn{2.85}{3} & 0.447 & \xpn{3.6}{-4} & 3.118 &  0.45 & 0.8586 & 
               2.108 & 1.662 & 3d \\
              &  1.0 & \xpn{6.35}{2} & 0.020 & \xpn{5.4}{-5} & 3.535 &  1.18 & 1.2690 & 
               2.113 & 1.328 & 3a \\
          5.0 &  0.4 & \xpn{1.78}{3} & 0.133 & \xpn{1.3}{-4} & 3.504 &  0.72 & 1.0449 & 
               2.260 & 1.243 & 3a \\
              &  0.6 & \xpn{7.92}{2} & 0.029 & \xpn{7.4}{-5} & 3.603 &  1.00 & 1.1419 & 
               2.259 & 1.318 & 3a \\
          6.4 &  0.2 & \xpn{1.34}{4} & 0.052 & \xpn{1.1}{-4} & 4.001 &  0.46 & 0.8992 & 
               2.626 & 1.534 & 2d \\

          \noalign{\smallskip}
	  \hline
	  \hline
         \end{tabular}
	 \end{center}
	\end{table*}
\setlength{\tabcolsep}{6pt}

We devide case BC mass transfer into two mass ranges, i.e. (i) 2.8 -- 3.2~\msun\ helium stars, in which a partially degenerate core develops at the end of the calculation, and mass transfer becomes dynamically unstable due to the development of a convective helium envelope, and (ii) 3.6 -- 6.4~\msun\ helium stars, in which the core is only weakly degenerate and a neon burning convective core develops at the end of the calculation. Typical examples for the two mass ranges are presented in Figs.~\ref{doublens:fig:2p8bc} -- \ref{doublens:fig:3p6bc}, and the overall results are presented in Table~\ref{doublens:tab:outcome}. Until the end of the calculation the neutron star does not accrete a significant amount of matter, its mass remains 1.4~\msun\ and therefore is not displayed in the table. Detailed discussions of the evolution of helium stars in binary systems have already been given in Paper~I, and therefore we will focus our discussion here on the late stages of the evolution.

\subsection{Roche-lobe overflow from 2.8 -- 3.2 $\bmath{\mathrm{M_{\sun}}}$ helium stars}
\label{doublens:subsec:low}

	\begin{figure}
  	 \centerline{\includegraphics[width=84mm]{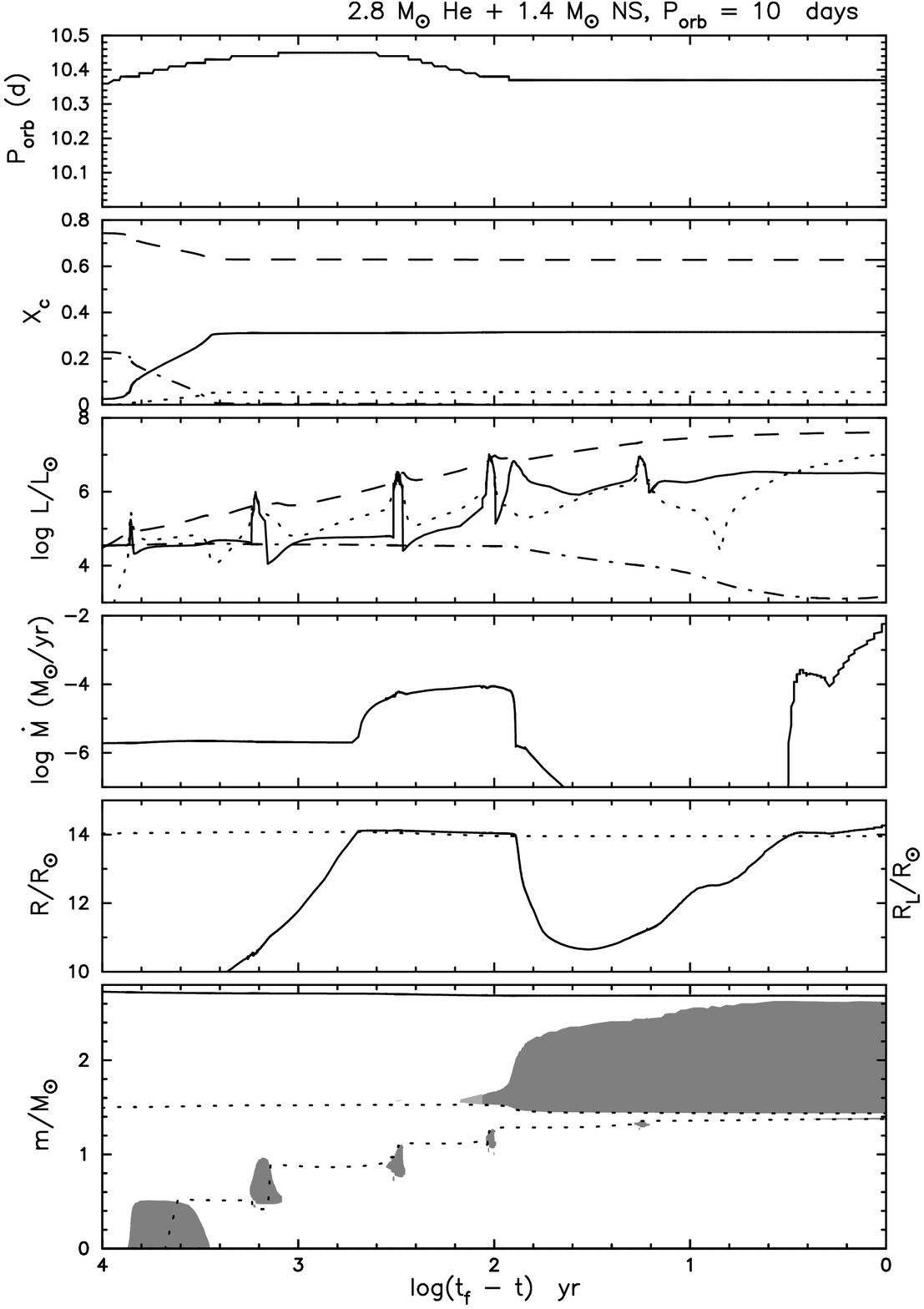}}
	 \caption[]{The evolution of a 2.8~\msun\ helium star with a 1.4~\msun\ 
                    neutron star companion with an orbital period of $10^{\mathrm{d}}$ 
                    starting from carbon core burning. The orbital evolution is
                    presented in the first panel. The second panel shows the central 
                    abundances: dashed-dotted-, dashed-, solid, and dotted-lines are C, 
                    O, Ne, and Mg abundances, respectively. The third panel gives the 
                    stellar luminosity (dash-dotted-), and the contributions of helium 
                    burning (solid-), carbon burning (dotted-), and neutrino losses 
                    (dashed-line). The fourth panel presents the mass-loss rate; while 
                    the fifth panel the stellar (solid-) and Roche (dotted-line) radii 
                    in solar units. The sixth panel shows the evolution of the stellar 
                    interior. The upper and lower dotted lines are the CO and ONeMg core 
                    masses, respectively, and the dark- and light-shaded areas mark the 
                    convective and semiconvective burning regions.}
	 \label{doublens:fig:2p8bc}
	\end{figure}

	\begin{figure}
 	 \centerline{\includegraphics[width=84mm]{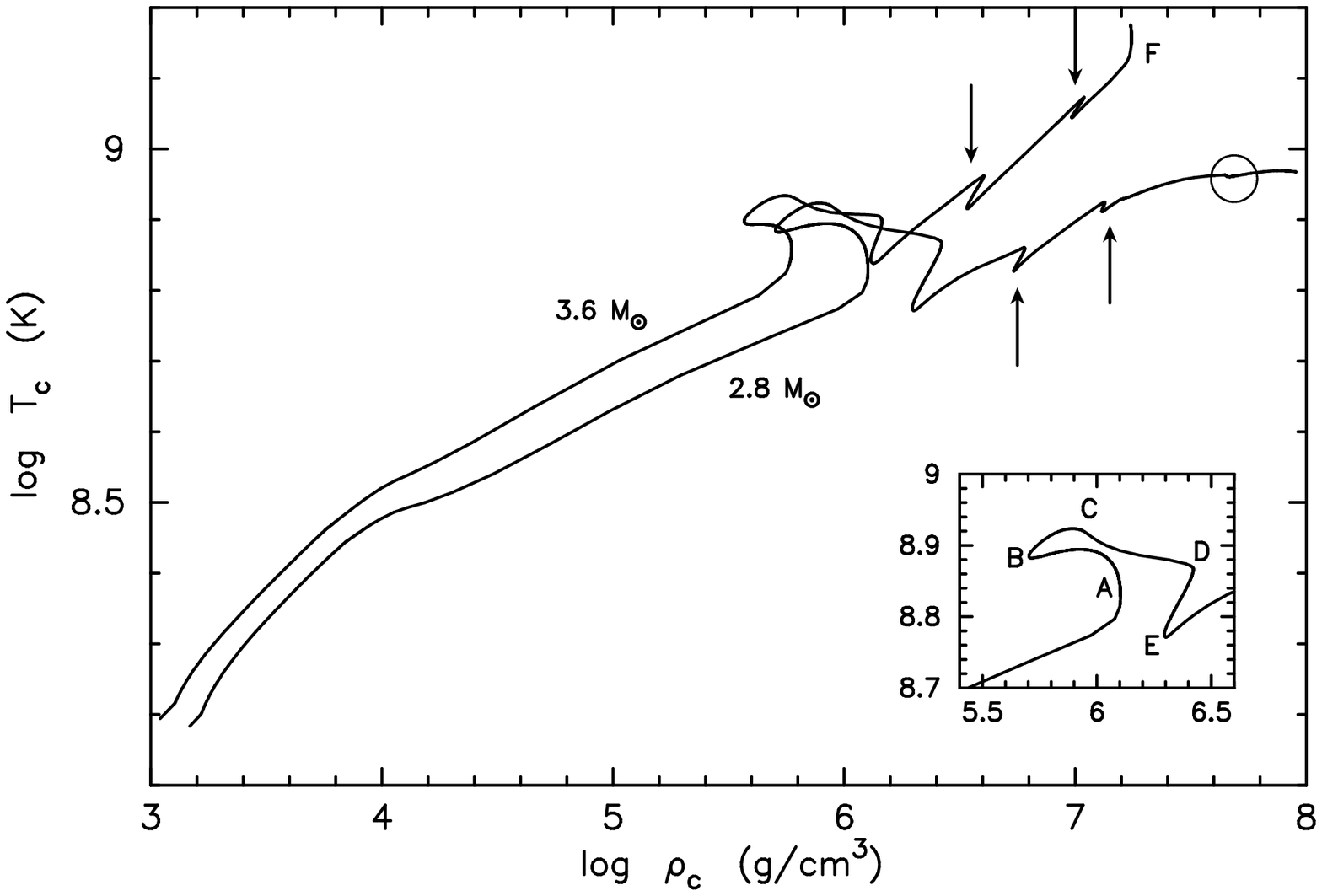}}
	 \caption[]{The evolutionary tracks in the central density-central 
                    temperature plane of a 2.8~\msun\ and a 3.6~\msun\ helium star
                    which go through case BC mass transfer. Point A corresponds to
                    carbon core burning, B (and C): the exhaustion (and depletion) 
                    of carbon in the core, D: the start of convective carbon shell burning, 
                    E: the end of convective carbon shell burning, and F: neon ignition. 
                    The arrows indicate the subsequent convective carbon shell burnings.}
	 \label{doublens:fig:rhotemp}
	\end{figure}

In all systems considered, mass transfer takes place on the thermal timescale. In contrast to case BB evolution, the mass-transfer rate is lower at larger initial period (see Table~\ref{doublens:tab:outcome}). This is probably because mass transfer is initiated at a very late stage and there is insufficient time to reach the full thermal-timescale rate. Fig.~\ref{doublens:fig:2p8bc} shows the evolution of a 2.8~\msun\ helium star with a neutron star in a $10^{\mathrm{d}}$ orbital period. A convective shell develops in the helium envelope before the third carbon-burning convective shell appears. The helium-burning convective shell causes a decrease of the mass of the CO core, $M_{\mathrm{CO}}$ (defined as the central mass with helium abundance less than 0.1, which coincides with the lower limit of the helium-burning convective shell in the envelope). The convective shell penetrates outward both in mass and in radius (at the end of calculation, the upper boundary of the convective shell is located at 98 per cent of the total mass and 54 per cent of the total radius of the star). Because the mass of the envelope is dominated by the convective shell, the helium envelope behaves as a convective envelope. As a result of mass loss from a star with a convective envelope, the radius does not shrink as in the case of a radiative envelope. This response of the star drives an enormous increase in \mdot, as can be seen in Fig.~\ref{doublens:fig:2p8bc}. Mass transfer appears to become dynamically unstable which would lead to a CE and spiral-in phase.

The start of each convective carbon burning phase (both in the core and in subsequent shells) causes the core to expand. As a result of the mirror principle, the envelope tends to contract and the star shrinks. In between the convective phases, the core contracts and the envelope expands again. This is reflected in the behaviour of the mass-transfer rate, which is sometimes interupted by a detached phase. In systems with closer orbits, a detached phase does not always occur (although the behaviour of expansion and contraction of the star is also noticed). The helium envelope in these systems is thinner such that the outward penetration of the helium-burning convective shell takes place relatively fast. However, whether a detached phase occurs or not does not affect the final situation, since all systems in this mass range are expected to end up in a CE phase.

The evolution of the central density and temperature of 2.8 and 3.6~\msun\ helium stars is presented in Fig.~\ref{doublens:fig:rhotemp}. At the end of the evolution of a 2.8~\msun\ helium star, the ONeMg core becomes partially degenerate. As a consequence of neutrino losses, $T_{\mathrm{c}}$ remains relatively constant as the core becomes denser at the onset of the fourth consecutive carbon shell burning (marked by a circle in Fig.~\ref{doublens:fig:rhotemp}), while the shell with maximum temperature moves outward. We will discuss the final stage of evolution of helium stars in this range of mass in Sect.~\ref{doublens:subsec:collapse}.

\subsection{Roche-lobe overflow from 3.6 -- 6.4 $\bmath{\mathrm{M_{\sun}}}$ helium stars}
\label{doublens:subsec:high}

	\begin{figure}
 	 \centerline{\includegraphics[width=84mm]{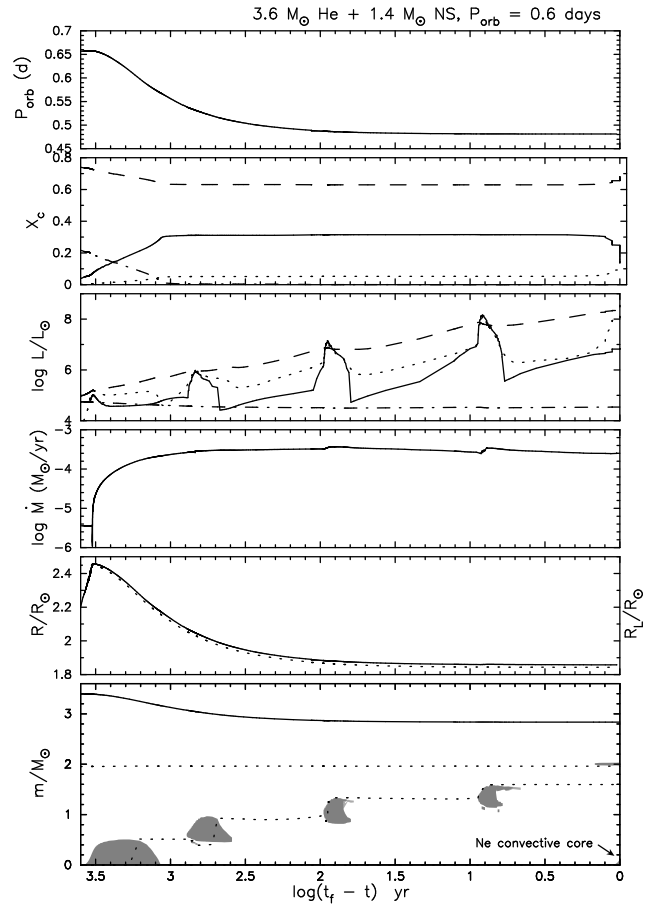}}
	 \caption[]{Same as Fig.~\ref{doublens:fig:2p8bc} for a 3.6~\msun\ helium star
                    with an orbital period of $0\fd6$.}
	 \label{doublens:fig:3p6bc}
	\end{figure}

Fig.~\ref{doublens:fig:3p6bc} shows the evolution of a 3.6~\msun\ helium star with a neutron star in a $0\fd6$ orbital period. A convective shell also appears in the helium envelope of stars in this range of mass. However, we do not find that the shell penetrates outward nor that \mdot\ increases enormously like in the lower-mass helium stars (see the fourth and sixth panels of Fig.~\ref{doublens:fig:3p6bc}). Although we could not follow the evolution further than neon ignition, we expect that a CE phase probably does not occur in this range of mass.

\begin{figure}
 	 \centerline{\includegraphics[width=84mm,angle=270.]{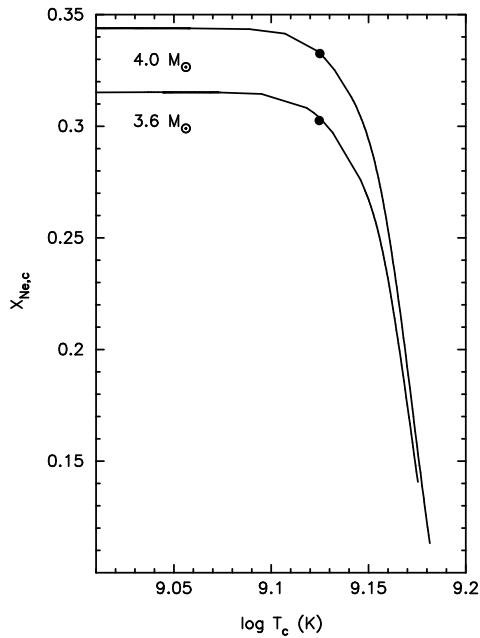}}
	 \caption[]{The central temperature and central neon abundance in 
                    3.6~\msun\ and 4.0~\msun\ helium stars around the moment 
                    of neon ignition. Solid circles mark the ignition of neon 
                    in the convective core.}
	 \label{doublens:fig:neon}
	\end{figure}

The core becomes only weakly degenerate. Before the third carbon-burning convective shell appears, the shell with maximum temperature moves to the centre. When the central temperature is $\sim$ \xpn{1.3}{9}~{K}, neon burning occurs in a convective core. Before the convective ignition, the central neon abundance has decreased by $\sim$ 3 per cent by radiative burning (see Fig.~\ref{doublens:fig:neon}). Convective neon core burning is found only in the 3.6~\msun\ model with orbital period of $0\fd6$ and the 4.0~\msun\ model with orbital period of $0\fd5$. In other models, the decrease in neon abundance is also noticed, but at the end of our calculations the central temperature is not high enough yet for convective neon ignition. The onset of convective neon ignition is represented by point F in Fig.~\ref{doublens:fig:rhotemp}. The evolutionary track at this point is similar to point A, i.e. the appearance of the convective core and the rapid increase of the neon burning rate cause a decrease in density while the temperature increases. The mass of the ONeMg core at the ignition of neon is 1.6~\msun\ (in a 3.6~\msun) -- 1.7~\msun\ (in a 4.0~\msun\ helium star). Here the ONeMg core is defined as the central mass that contains less than 10 per cent carbon, which coincides with the upper limit of the carbon burning convective shell.

\section{The possible remnants of case BB and case BC evolution}
\label{doublens:sec:remnants}

\subsection{The late stages of case BC mass transfer and comparison with case BB evolution}
\label{doublens:subsec:caseBB}

We have discussed in Sect.~\ref{doublens:subsec:low} that a convective envelope develops in helium stars of 2.8 -- 3.2~\msun\ which go through case BC evolution, and therefore they will undergo a CE phase at the end of their evolution. Such a convective shell in the helium envelope was also found in some of our case BB calculations (Paper~I). We could not always follow the evolution far enough to see if outward penetration of this shell does take place. However, we found that the penetration and enormous increase in \mdot\ occur in helium stars of 2.8 -- 3.2~\msun\ with various initial periods of case BB evolution. Therefore, we suggest that a CE phase is a typical characteristic of the late stage of evolution of helium star-neutron star binaries in this range of mass.

In helium stars more massive than 3.2~\msun, the outward penetration of the helium-burning convective shell is only found in 3.6 -- 3.8~\msun\ stars with very close orbits ($P_{\mathrm{i}} \la 0\fd25$, i.e. case BB mass transfer)\footnote{In Paper~I, it was mentioned that the penetration of the helium-burning convective shell into the envelope takes place in helium stars of various masses (cf. table~4). In some cases the appearance of the convective burning shell was noticed, and it was assumed that once the shell appears, it will penetrate outward. Later while working on this paper, we found that the appearance of a convective burning shell in the helium envelope is not always followed by an outward penetration.}. We could not draw a conclusion about the 3.4~\msun\ models because all calculations for this mass stopped before the appearance of the convective shell in the helium envelope, but they probably behave in a similar way. Assuming that a penetration of the convective shell in the helium envelope accompanied by an enormous increase in \mdot\ is the indication of the start of dynamically unstable mass transfer, we suggest that a CE phase also occurs in 3.2 -- 3.8~\msun\ helium stars in very close orbits. Wider-orbit systems in this range of mass as well as helium stars more massive than 3.8~\msun\ probably will not go through a CE phase. 

A similar study of the evolution of helium star-neutron star binaries has recently been carried out by Ivanova et al. (2002). Although the latter work is similar to ours, they come to different conclusions. They do not find that lower-mass helium stars develop a convective envelope at the end of their evolution, even though their calculations were purportedly done up to a more advanced stage of evolution (i.e. up to oxygen burning) than our calculations, which extend up to radiative neon ignition. On the other hand, they find that helium stars with masses between 3.3 and 5~\msun\ in orbital periods less than $0\fd3$ (which they called case CEB) undergo stable mass transfer, but the mass-transfer rate exceeds a critical value which is a function of the orbital period, such that a CE may form which probably leads to a merger. However, it is difficult to compare our models with their calculations as they do not provide the detailed interior structure of their models.

\subsection{Final stages of evolution of 2.8 -- 3.2 $\bmath{\mathrm{M_{\sun}}}$ helium stars}
\label{doublens:subsec:collapse}

	\begin{figure*}
 	 \centerline{\includegraphics[width=100mm,angle=270.]{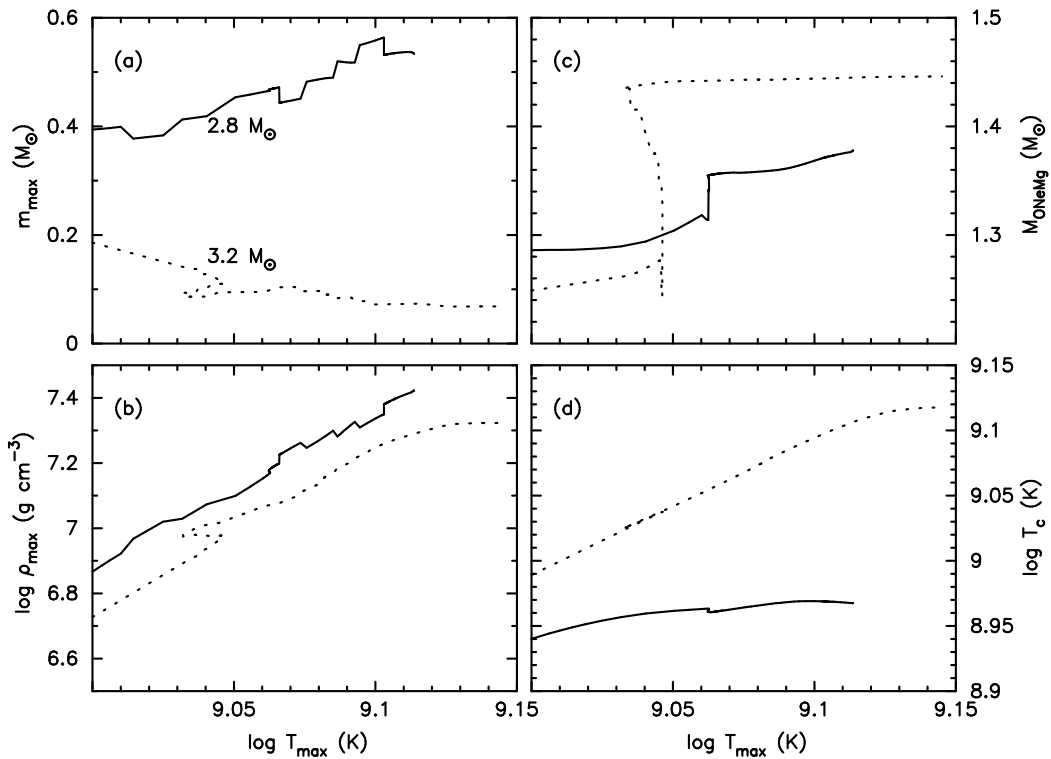}}
	 \caption[]{The location of the shell with maximum temperature ({\it a}), 
                    the density of this shell ({\it b}); the corresponding mass 
                    of the ONeMg core ({\it c}), and the central temperature 
                    ({\it d}) for 2.8~\msun\ (solid line) and 3.2~\msun\ (dotted 
                    line) helium stars, presented as a function of maximum 
                    temperature.}
	 \label{doublens:fig:tmax}
	\end{figure*}

We have discussed in Sect.~\ref{doublens:subsec:high} that before 3.6~\msun\ and 4.0~\msun\ helium stars ignite neon convectively in the centre, the central neon abundance decreases by $\sim$ 3 per cent. The decrease of the neon abundance in the shell with maximum temperature is also found in 2.8 -- 3.2~\msun\ helium stars. Fig.~\ref{doublens:fig:tmax} compares the conditions in the centre and in the shell with maximum temperature for 2.8~\msun\ and 3.2~\msun\ stars. The maximum temperature increases with time. In 2.8 -- 2.9~\msun\ helium stars, the shell with maximum temperature moves outward (see panel ({\it a}) in Fig.~\ref{doublens:fig:tmax}) and at the end of the calculation the temperature is still lower than \xpn{1.3}{9}~{K}, i.e. insufficient to enable convective neon burning. In a 3.2~\msun\ helium star, the shell moves inward and at the end of the calculation the maximum temperature reaches \xpn{1.4}{9}~{K}, but an off-centre neon burning convective shell has not yet developed. Panel ({\it c}) of Fig.~\ref{doublens:fig:tmax} shows that in both cases, the mass of the ONeMg core at the end of the calculation is large enough to ignite neon, i.e. higher than 1.37~\msun\ (Nomoto 1984). Hence, we argue that at the end of our 2.8 -- 3.2~\msun\ calculations the stars are close to off-centre neon ignition. For comparison, Nomoto \& Hashimoto (1988) found that a 3.3~\msun\ helium core ignites neon in the centre while 2.8 -- 3.2~\msun\ helium cores ignite neon off-centre. They conclude that 3.3~\msun\ is the critical mass between off-centre and central ignition of neon.

After off-centre neon ignition, the future of a partially-degenerate ONeMg core depends on whether the burning shell is able to reach the centre (Nomoto \& Hashimoto 1988). If it does reach the centre, then subsequent nuclear burning stages of O and Si will form an iron core, which will collapse due to photodisintegration. Otherwise, part of the ONeMg core is left unburned and becomes highly degenerate, and the core will eventually collapse due to electron captures. The density at the burning front also plays an important role in determining the future of the core. If the density is higher than $10^{8} \, {\mathrm{g \, cm^{-3}}}$, neon shell burning becomes so explosive that a dynamical event, such as the ejection of the helium envelope, may occur. The latter phenomenon would have interesting consequences for the binary systems of our interest. The ejection of the helium layer would leave a bare CO core as in the case of a spiral-in phase (Sect.~\ref{doublens:subsec:supernova}), yielding a lower pre-SN mass. 

Fig.~\ref{doublens:fig:tmax} shows that both the maximum and central temperature of the 3.2~\msun\ model are already close to that required for neon burning (panel ({\it d})). The shell with maximum temperature tends to move inward, and most probably will reach the centre. The density at the neon burning front is lower than $10^{8} \, {\mathrm{g \, cm^{-3}}}$ (panel ({\it b})). Hence, a 3.2~\msun\ helium star is expected to undergo core collapse due to photodisintegration, and most probably a dynamical event will not occur. The future of the 2.8~\msun\ model is less clear. The shell with maximum temperature moves outward which might leave part of the ONeMg core unburned. The central density is close to $10^{8} \, {\mathrm{g \, cm^{-3}}}$. Whether a dynamical event occurs or not depends on how close the neon burning front is from the centre. Nomoto \& Hashimoto (1988) argue that in a 2.8~\msun\ helium core, as a result of electron captures in the neon burning shell, the mean molecular weight $\mu_{\mathrm{e}}$ would increase above 2 which implies a lower Chandrasekhar mass. Therefore the gravothermal specific heat of the core remains negative and the central temperature will continue to increase. Neon burning will reach the centre and photodisintegration will trigger the core collapse.

\subsection{Spiral-in phase vs. supernova explosion}
\label{doublens:subsec:supernova}

Helium stars more massive than 3.3~\msun\ do not undergo a CE and spiral-in phase in their late stage of evolution. The further evolution is straightforward. The core is weakly degenerate, and will collapse due to photodisintegration; the mass and period prior to the explosion are the same as those at the end of RLOF (i.e. $M_{\mathrm{t}} = M_{\mathrm{o}}$, $P_{\mathrm{t}} = P_{\mathrm{o}}$).

Helium stars of 2.8 -- 3.3~\msun\ and those less massive than 3.8~\msun\ in very close orbits undergo a CE phase at the end of their evolution. The first question we will try to answer is whether the system will survive CE evolution, and how long the spiral-in phase will last. We ignore the possibility that the stars might experience a dynamical event due to the high density at the neon burning shell. Neon burning (or the increase of temperature close to that required for neon ignition) and the high $L_{\nu}$ suggest that the helium star is already close to core collapse. The next question is whether the SN explosion takes place before or after the neutron star completes the spiraling-in process in the envelope of the helium star. This depends on the competition between the spiral-in timescale and the remaining time until the explosion. 

\subsubsection{The decay timescale of the spiral-in phase}
\label{doublens:subsubsec:decay}

For a CE situation where the accretor is significantly less massive than the donor, so that the accretor does not cause a considerable perturbation to the structure of the donor, the orbital decay can be expressed as the change in orbital energy due to the drag force (Bondi \& Hoyle 1944, Shima et al. 1985), i.e.
	\begin{eqnarray}
          - \frac{G \, M_{a} \, M_{\mathrm{NS}}}{2 \, a^{2}} \frac{da}{dt} = 
          \xi(\mu) \, \pi \, R_{\mathrm{A}}^{2} \, \rho_{a} \, V^{3} 
	  \label{doublens:eq:decay}
	\end{eqnarray}
where $M_{a}$ is the mass in the giant interior to radius $a$, $\rho_{a}$ is the local density at separation $a$, and $V$ is the relative velocity between the secondary and the CE. $\mu$ is the Mach number, i.e. $\mu = V / V_{\mathrm{s}}$ where $V_{\mathrm{s}}$ is the speed of sound. The accretion radius $R_{\mathrm{A}}$ is approximated by
	\begin{eqnarray}
          R_{\mathrm{A}} = \frac{2 \, G \, M_{\mathrm{NS}}}{V^{2} + V_{s}^{2}}
	  \label{doublens:eq:accretion}
	\end{eqnarray}
The function $\xi(\mu)$ determines the dissipation rate, and is of order 2 -- 4 in the supersonic case (Bondi \& Hoyle 1944, Shima et al. 1985). We applied $\xi(\mu) = 2.5$ which is the value for a steady-state set up after a violent perturbation (Bondi \& Hoyle 1944).

In order to estimate the timescale of the CE and spiral-in phase, we define the decay timescale as $\tau_{\mathrm{decay}} = a / \dot{a}$ and calculate the ratio of the decay timescale to the Keplerian timescale which is expressed as (Livio \& Soker 1988 and references therein):
	\begin{eqnarray}
          \beta_{\mathrm{CE}} & \equiv &  
          \frac{\tau_{\mathrm{decay}}}{\tau_{\mathrm{Kep}}} \nonumber \\
                              &   =    &
          \frac{1}{12 \, \pi} \, G(\mu) \,
          \left[ \frac{M_{a} + M_{\mathrm{NS}}}{M_{\mathrm{NS}}} \right]
          \left( \frac{V_{\mathrm{s}}}{V_{\mathrm{Kep}}} \right)
          \left( \frac{{\bar{\rho}}_{a}}{\rho_{a}} \right)
	  \label{doublens:eq:beta}
	\end{eqnarray}
where $V_{\mathrm{Kep}}$ is the Keplerian orbital velocity, and ${\bar{\rho}}_{a}$ is the average density in the donor interior to radius $a$. $G(\mu)$ is a function of the Mach number $\mu$ given by
	\begin{eqnarray}
          G(\mu) = \frac{1}{\xi(\mu)} 
          \frac{(\mu^{2} + 1)^{2}}{\mu^{3}}
	  \label{doublens:eq:mach}
	\end{eqnarray}
The parameter $\beta_{\mathrm{CE}}$ measures the importance of local (three-dimensional) effects in the spiraling-in process. If $\beta_{\mathrm{CE}} \la 1$, energy is deposited locally and neither spherical nor cylindrical symmetry can be assumed.

As a result of the deposition of orbital angular momentum into the envelope, the envelope is spun up, and the relative velocity between the neutron star and the envelope is reduced. The drag force decreases, prolonging the spiraling-in process. We calculate the ratio of the spin-up timescale of the envelope to the decay timescale as (Livio \& Soker 1988)
	\begin{eqnarray}
          \gamma_{\mathrm{CE}} & \equiv & 
          \frac{\tau_{\mathrm{spin-up}}}{\tau_{\mathrm{decay}}} \nonumber \\
                               &   =    &
          1.2 \, \mu^{2} 
          \left[ \frac{M_{a} + M_{\mathrm{NS}}}{M_{\mathrm{NS}}} \right]
          \left( \frac{{\tilde{\rho}}_{a}}{{\bar{\rho}}_{a}} \right)
          \left( \frac{V_{\mathrm{s}}}{V_{\mathrm{Kep}}} \right)^{2}
	  \label{doublens:eq:gamma}
	\end{eqnarray}
where
	\begin{eqnarray}
          {\tilde{\rho}}_{a} = \frac{5}{a^{5}} \int_{R_{\mathrm{in}}}^{a} r^{4} \rho(r) dr
	  \label{doublens:eq:tilderho}
	\end{eqnarray}
$R_{\mathrm{in}}$ is the radius at the giant's core-envelope boundary. A considerable spin-up of the envelope is expected to occur if $\gamma_{\mathrm{CE}} \la 1$.

	\begin{figure}
 	 \centerline{\includegraphics[width=84mm]{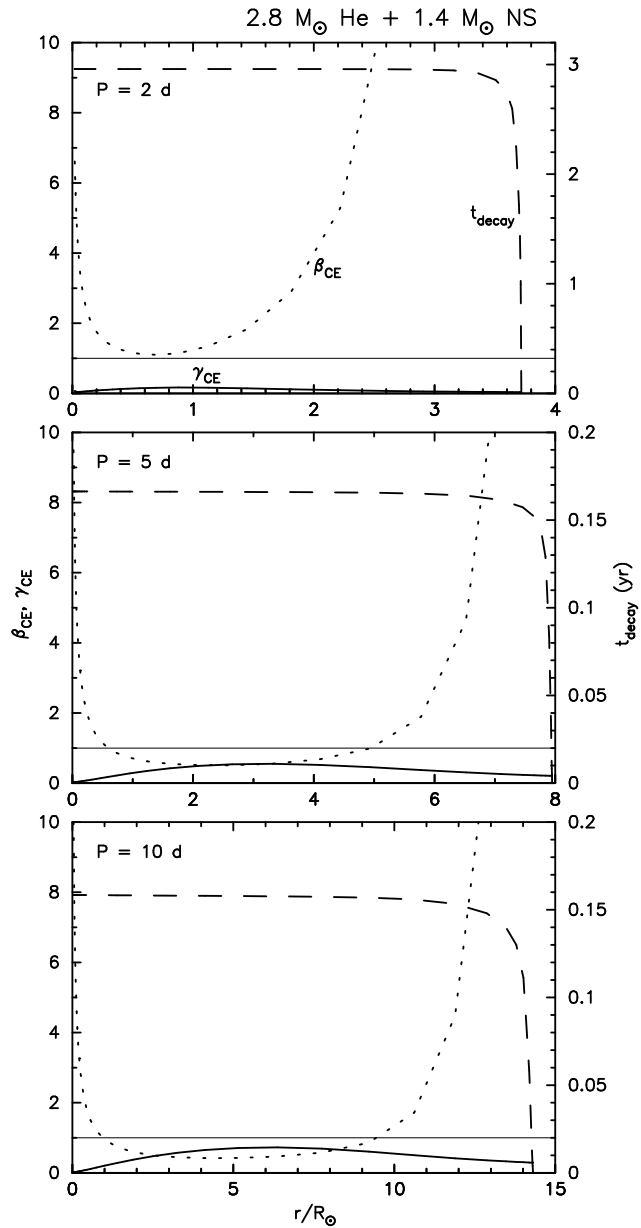}}
	 \caption[]{The physical parameters in the common envelope of a 2.8~\msun\ 
                    helium star with a 1.4~\msun\ neutron-star companion at the 
                    last model of our calculations for three different initial periods: 
                    the decay timescale with the scale on the right y-axis (dashed-), 
                    and the parameters $\beta_{\mathrm{CE}}$ (dotted-) and 
                    $\gamma_{\mathrm{CE}}$ (solid-line).}
	 \label{doublens:fig:decay}
	\end{figure}

The above estimation of the decay timescale is only valid if the donor is much more massive than the neutron star (cf. e.g. Iben \& Livio 1993). In our calculation we have $M_{\mathrm{NS}} \sim M_{\mathrm{o}}$, and hence the above derivation is not completely valid, but nevertheless we use it to obtain a rough estimate of the decay timescale.

Fig.~\ref{doublens:fig:decay} presents the $\beta_{\mathrm{CE}}$ and $\gamma_{\mathrm{CE}}$ parameters as a function of radius in the envelope of a 2.8~\msun\ helium star with three different initial periods in the final models of our calculations. Here the relative velocity between the neutron star and the envelope of the helium star is assumed to be Keplerian. In all cases we find $\gamma_{\mathrm{CE}} < 1$, suggesting that spin-up of the envelope takes place at relatively early stages, and therefore the spiraling-in phase proceeds on a timescale longer than the estimated decay timescale. The $\gamma_{\mathrm{CE}}$ parameter increases and local effects become more important ($\beta_{\mathrm{CE}}$ decreases) with increasing initial period. This can be understood because a helium star in a wider system has more or less the same core mass as that in a closer system, but with a more extended envelope and higher density gradient, and therefore is more centrally condensed (smaller $\bar{\rho}_{a}$ but larger $\bar{\rho}_{a} / \rho_{a}$).

By integrating eq.~(\ref{doublens:eq:decay}) over the envelope of the helium star, we estimate the timescale of the spiral-in phase, $t_{\mathrm{decay}}$, as presented in Fig.~\ref{doublens:fig:decay}. Note that $t_{\mathrm{decay}} \neq \tau_{\mathrm{decay}}$ as defined above; $\tau_{\mathrm{decay}}$ is the local decay timescale while $t_{\mathrm{decay}}$ represents the time required to spiral down to radius $r$. Taking these estimates at face value suggests that the spiraling-in process in the system with $P_{\mathrm{i}} = 10^{\mathrm{d}}$ will last for 0.16~yr, while in the system with $P_{\mathrm{i}} = 2^{\mathrm{d}}$ it lasts for 3~yr. However, because $\gamma_{\mathrm{CE}} < 1$ for all systems, in particular for $P_{\mathrm{i}} = 2^{\mathrm{d}}$, the actual spiral-in timescale will be longer.

 Furthermore, we have assumed that the structure of the envelope does not change during the spiral-in. If we consider that the deposition of orbital energy actually causes the envelope to expand, then the expansion causes a decrease in $\rho_{a} / {\bar{\rho}}_{a}$. This will lead to a slower orbital decay, by eq.~(\ref{doublens:eq:decay}). The quantity ${\tilde{\rho}}_{a}$, which depends only on the structure of the envelope, decreases more significantly than ${\bar{\rho}}_{a}$ (which is determined also by the core density) as the result of this expansion. Accordingly, $\gamma_{\mathrm{CE}}$ decreases, and the spiral-in proceeds even more slowly. Therefore, the decay timescale we have estimated above should be regarded as the lower limit. Note that Podsiadlowski (2001) has included the envelope expansion in his CE calculation and found that the spiral-in is initially very rapid and slows down after significant envelope expansion has taken place.

\subsubsection{The time left until the supernova explosion}
\label{doublens:subsubsec:explosion}

We first consider the possibility that the neon-burning shell does not reach the centre and therefore core collapse is triggered by electron captures. Because the core loses energy mainly in neutrinos, we can estimate the remaining time until the explosion by comparing the change in binding energy to the neutrino luminosity, $L_{\nu}$. We calculate the binding energy of the core of the last model, $E_{\mathrm{o}}$, and estimate the binding energy at the onset of electron captures, $E_{\mathrm{f}}$. By assuming that at this point the core is completely degenerate, we solve Chandrasekhar's differential equation for the structure of a white dwarf (Kippenhahn \& Weigert 1994). We applied $\rho_{\mathrm{c}}$ = \xpn{3.7}{9}~$\mathrm{g \, cm^{-3}}$ as the central density at the onset of electron capture (Nomoto 1987, Miyaji et al. 1980, Miyaji \& Nomoto 1987). By assuming that $L_{\nu}$ remains constant after the last model, we estimate the time until the explosion as $\Delta E / L_{\nu}$, where $\Delta E = E_{\mathrm{f}} - E_{\mathrm{o}}$. Considering that $L_{\nu}$ probably increases instead of remaining constant, the timescale we derive gives an upper limit. For the 2.8~\msun\ helium star, we find that the maximum time is 70~yr for the system with $P_{\mathrm{i}} = 10^{\mathrm{d}}$ and 103~yr for the system with $P_{\mathrm{i}} = 2^{\mathrm{d}}$. For a 3.2~\msun\ helium star, the maximum time is 28~yr.

Next we consider the possibility that the neon burning shell does reach the centre, and core collapse is triggered by photodisintegration which seems more likely than electron capture given the discussion in Sec.~\ref{doublens:subsec:collapse}. A comparison with a detailed evolution calculation (Heger 2002, private communication) yields that a 1.696~\msun\ CO core will undergo core collapse $\sim$ 20 yr after off-centre neon ignition. This core mass is about the same as the CO core mass of our 3.2~\msun\ model. Our 2.8~\msun\ helium star has a CO core of 1.45~\msun, and probably needs a longer time before undergoing core collapse. We conclude, therefore, that the 3.2~\msun\ helium star will undergo core collapse in 20 -- 30~yr, and the 2.8~\msun\ model needs 20 -- 100~yr before it collapses.

Although we have tried to estimate the timescales for the spiral-in phase and for the star to undergo core collapse, the exact timescales remain uncertain. Therefore, there are two open possibilities for the outcome. If the orbital decay timescale is shorter than the time until collapse, the helium star explodes after the spiral-in phase is terminated. If the time until collapse is shorter than the orbital decay timescale, the SN explosion takes place before the neutron star completes the spiraling-in process (i.e. inside the CE). This has important consequences for the pre-SN mass and period of the system.

\subsubsection{The possible remants of the lower mass helium stars}
\label{doublens:subsubsec:outcome}

	\begin{table*}  
         \caption[]{The remnants of case BB and case BC evolutions from 2.8 -- 3.2~\msun\ 
                    helium stars after surviving CE and spiral-in phase: the initial mass 
                    and period, the pre-CE mass and period, the mass and radius of the 
                    core, the binding energy parameter, the post-CE period and separation, 
                    the post-SN separation and eccentricity, and the merger timescale.}
	 \label{doublens:tab:post-CE}
	 \begin{center}
	 \begin{tabular}{ccc|cr|cccrc|ccc}
	  \hline
	  \hline
          \noalign{\smallskip}
          Case & $M_{\mathrm{i}}$ & $P_{\mathrm{i}}$ & $M_{\mathrm{o}}$ & $P_{\mathrm{o}}$~ & 
          $M_{\mathrm{core}}$ & $R_{\mathrm{core}}$ & $\lambda$ &
          $P_{\mathrm{t}}$~ & $a_{\mathrm{t}}$ & $a_{\mathrm{f}}$ & $e_{\mathrm{f}}$ & 
          $\tau_{\mathrm{merger}}$ \\
          & \msun & d & \msun & d~~ & \msun & $10^{-2}$~\rsun & & m~~ & \rsun & \rsun & &
          yr \\ 
          \noalign{\smallskip}
	  \hline
          \noalign{\smallskip}
          BB
          & 2.8 & 0.08 & 1.528 & 0.078 & 1.428 & 3.248 & 0.137 & 14.59 & 0.279 & 0.282 & 0.01 & \xpn{1.8}{5} \\
          &     & 0.5  & 1.686 & 0.467 & 1.506 & 3.464 & 0.055 & 12.76 & 0.257 & 0.267 & 0.04 & \xpn{1.4}{5} \\
          &     & 1    & 1.711 & 0.926 & 1.456 & 2.121 & 0.072 & 21.42 & 0.361 & 0.368 & 0.02 & \xpn{5.1}{5} \\
          & 2.9 & 0.08 & 1.593 & 0.074 & 1.480 & 3.431 & 0.145 & 12.89 & 0.258 & 0.266 & 0.03 & \xpn{1.4}{5} \\
          &     & 0.5  & 1.775 & 0.439 & 1.563 & 3.650 & 0.061 & 10.88 & 0.233 & 0.247 & 0.06 & \xpn{1.0}{5} \\
          &     & 1    & 1.834 & 0.868 & 1.511 & 2.522 & 0.052 &  8.93 & 0.203 & 0.211 & 0.04 & \xpn{5.5}{4} \\
          & 3.1 & 0.08 & 1.717 & 0.064 & 1.585 & 3.635 & 0.161 & 10.62 & 0.229 & 0.245 & 0.07 & \xpn{9.9}{4} \\
          &     & 0.3  & 1.887 & 0.233 & 1.580 & 2.574 & 0.109 &  7.45 & 0.181 & 0.193 & 0.06 & \xpn{3.8}{4} \\
          &     & 0.5  & 1.981 & 0.397 & 1.684 & 4.139 & 0.074 &  8.14 & 0.194 & 0.216 & 0.10 & \xpn{5.8}{4} \\
          & 3.2 & 0.08 & 1.786 & 0.060 & 1.641 & 3.778 & 0.171 &  9.62 & 0.216 & 0.236 & 0.09 & \xpn{8.5}{4} \\
          &     & 0.3  & 1.977 & 0.219 & 1.616 & 2.737 & 0.125 &  6.65 & 0.169 & 0.183 & 0.08 & \xpn{3.1}{4} \\
          &     & 0.5  & 2.080 & 0.379 & 1.742 & 4.376 & 0.080 &  7.18 & 0.180 & 0.205 & 0.12 & \xpn{4.7}{4} \\
          & 3.4 & 0.08 & 1.919 & 0.054 & 1.745 & 4.058 & 0.191 &  8.06 & 0.194 & 0.221 & 0.12 & \xpn{6.3}{4} \\
          &     & 0.1  & 1.961 & 0.066 & 1.771 & 4.197 & 0.177 &  8.11 & 0.196 & 0.226 & 0.13 & \xpn{6.8}{4} \\
          & 3.6 & 0.09 & 2.059 & 0.057 & 1.754 & 2.872 & 0.203 &  4.32 & 0.128 & 0.147 & 0.13 & \xpn{1.2}{4} \\
          &     & 0.25 & 2.258 & 0.158 & 1.935 & 4.785 & 0.125 &  6.18 & 0.166 & 0.205 & 0.19 & \xpn{4.3}{4} \\
          & 3.7 & 0.09 & 2.124 & 0.055 & 1.905 & 4.433 & 0.201 &  6.71 & 0.175 & 0.214 & 0.18 & \xpn{5.1}{4} \\
          &     & 0.2  & 2.301 & 0.126 & 1.956 & 5.008 & 0.139 &  5.20 & 0.148 & 0.185 & 0.20 & \xpn{2.8}{4} \\
          & 3.8 & 0.09 & 2.183 & 0.056 & 1.960 & 5.325 & 0.226 &  7.12 & 0.183 & 0.229 & 0.20 & \xpn{6.6}{4} \\
          &     & 0.25 & 2.366 & 0.159 & 1.877 & 3.416 & 0.139 &  3.79 & 0.119 & 0.143 & 0.17 & \xpn{1.1}{4} \\
          \noalign{\smallskip}
	  \hline
          \noalign{\smallskip}
          BC
          & 2.8 &  2 & 1.758 &  1.811 & 1.438 & 1.807 & 0.062 & 23.87 & 0.387 & 0.392 & 0.01 & \xpn{6.6}{5} \\
          &     &  5 & 2.387 &  4.660 & 1.433 & 1.707 & 0.112 & 22.33 & 0.370 & 0.374 & 0.01 & \xpn{5.5}{5} \\
          &     & 10 & 2.684 & 10.370 & 1.436 & 1.498 & 0.120 & 33.62 & 0.486 & 0.492 & 0.01 & \xpn{1.6}{6} \\
          & 2.9 &  2 & 2.043 &  1.710 & 1.469 & 1.928 & 0.099 & 16.88 & 0.308 & 0.316 & 0.03 & \xpn{2.8}{5} \\
          &     &  4 & 2.658 &  3.920 & 1.480 & 2.052 & 0.122 & 14.83 & 0.283 & 0.291 & 0.03 & \xpn{2.0}{5} \\
          &     &  6 & 2.791 &  6.267 & 1.504 & 2.811 & 0.050 &  5.55 & 0.148 & 0.154 & 0.04 & \xpn{1.5}{4} \\
          & 3.2 &  1 & 2.520 &  0.819 & 1.639 & 2.408 & 0.124 &  5.75 & 0.153 & 0.167 & 0.09 & \xpn{2.1}{4} \\
          &     &  2 & 3.020 &  2.047 & 1.626 & 2.395 & 0.111 &  5.34 & 0.146 & 0.159 & 0.08 & \xpn{1.7}{4} \\
          \noalign{\smallskip}
	  \hline
	  \hline
         \end{tabular}
	 \end{center}
	\end{table*}

Assuming that there is enough time to spiral in before the explosion, we will now investigate whether the system survives the CE and spiral-in phase. With the energy equation for CE evolution (Webbink 1984, de Kool 1990) 
	\begin{eqnarray}
	  \frac {a_{\mathrm{t}}} {a_{\mathrm{o}}} & = & 
	  \frac {M_{\mathrm{core}} \, M_{\mathrm{NS}}} {M_{\mathrm{o}}} \,\,
	  \frac {1} {M_{\mathrm{NS}} + 2 M_{\mathrm{env}}/
	  (\eta_{\mathrm{CE}}\, \lambda \, r_{\mathrm{L}})} 
	  \label{doublens:eq:webbink}
	\end{eqnarray}
we calculated the post-CE separation, $a_{\mathrm{t}}$. Here $M_{\mathrm{core}}$ and $M_{\mathrm{env}}$ are the masses of the core and envelope, respectively. $r_{\mathrm{L}} = R_{\mathrm{L}} / a$ is the dimensionless Roche radius of the helium star, and $a_{\mathrm{o}}$ is the pre-CE separation. $\eta_{\mathrm{CE}}$, the so-called efficiency parameter of CE, is taken to be 1. The parameter $\lambda$ describing the binding energy of the envelope to the core -- approximated as in Dewi \& Tauris (2000) -- is calculated from the bottom of the convective helium envelope taking into account the gravitational binding energy only (if the internal energy is also taken into account, $\lambda$ and therefore $a_{\mathrm{t}}$ would be larger). We find $\lambda$ is in the range 0.05 -- 0.12 (see Table~\ref{doublens:tab:post-CE}). We assume that the binary will survive the CE and spiral-in phase if the CO core does not immediately fill its Roche lobe in its new orbit. 

Assuming that the neutron star completes the spiral-in phase, the post-CE (i.e pre-SN) mass and period are presented in Table~\ref{doublens:tab:post-CE}. We find that all systems have final separations larger than the radius of the core ($a_{\mathrm{t}} \ga 3 \, R_{\mathrm{core}}$), suggesting that they all survive the CE and spiral-in phase. Even if we consider that the core could expand by a factor of 2 after the envelope is peeled off, in most cases it still would not fill its Roche lobe. However, we should realize that the product $\eta_{\mathrm{CE}} \, \lambda$ in eq.~(\ref{doublens:eq:webbink}) depends on the details of CE phase which are very uncertain and therefore the results in Table~\ref{doublens:tab:post-CE} give only approximate final separations. We find that the pre-SN orbit has a period of $\sim 0\fd01$. In such a tight orbit, the neutron star moves with a very large orbital velocity ($\sim \, 10^{3}$~\kms). This velocity is higher than a plausible kick velocity of a few hundreds \kms. Therefore the effect of a supernova kick is not as strong as in the case of systems in wider orbits. However, although an asymmetric explosion does not change the separation significantly, it can increase the eccentricity significantly, which in turn reduces the merger timescale (Belczynski, Bulik \& Kalogera, 2002). If we assume that the explosion is symmetric and that the SN remnant has a mass of 1.4~\msun, then the mass which leaves the system is less than half of the initial total mass, and therefore all systems will remain bound (Blaauw 1961).

After the explosion of the helium star, which is assumed to leave a neutron-star remnant (with mass $M_{\mathrm{NS,2}}$), the orbital evolution of the two neutron stars will be governed by gravitational-wave radiation as (Peters 1964)
	\begin{eqnarray}
          \frac{da}{dt} = - \frac{64}{5} \,\, \frac{G^{3}}{c^{5}} \,\,
          \frac{M_{\mathrm{NS}} \, M_{\mathrm{NS,2}} \, M_{\mathrm{T}}}
               {a^{3} \, (1 - e^{2})^{7/2}} \,\,
          \left( 1 + \frac{73}{24} \, e^{2} + \frac{37}{96} \, e^{4} \right)
	  \label{doublens:eq:gwr-sep}
	\end{eqnarray}
	\begin{eqnarray}
          \frac{de}{dt} = - \frac{304}{15} \,\, e \,\, \frac{G^{3}}{c^{5}} \,\,
          \frac{M_{\mathrm{NS}} \, M_{\mathrm{NS,2}} \, M_{\mathrm{T}}}
               {a^{4} \, (1 - e^{2})^{5/2}} \,\,
          \left( 1 + \frac{121}{304} \, e^{2} \right)
	  \label{doublens:eq:gwr-ecc}
	\end{eqnarray}
Assuming a symmetric explosion, we calculate the post-SN separation and eccentricity as will be explained in Sect.~\ref{doublens:subsec:symmetry}. These parameters are used to determine the merger timescales of these systems, which is calculated by integrating eqs.~(\ref{doublens:eq:gwr-sep}) and (\ref{doublens:eq:gwr-ecc}) (for the complete equations please refer to Peters (1964)). This merger timescale is much shorter than that of the observed DNSs (see Table~\ref{doublens:tab:post-CE}), and also shorter than their characteristic age (i.e. $\sim$ 300~Myr and 100~Myr, respectively, for B1913+16). This makes the probability for such systems to be observed very small. 

The existence of very tight-orbit DNSs has been also proposed by Belczynski et al. (2002), with the assumption that a CE phase occurs if the helium star is more massive than the neutron star. We have demonstrated in this work that a CE phase does occur, but only if the helium star has a mass in a certain range, i.e. 2.8 -- 3.3~\msun\ and 3.3 -- 3.8~\msun\ in $P_{\mathrm{orb}} \la 0\fd25$. With very short merger timescales, these DNSs would increase the detection rate of gravitational-wave radiation. Another implication is that the merger would take place relatively close to the host galaxy. Since a merger of a compact binary has long been thought to be one of the sources of $\gamma$-ray bursts, the existence of tight-orbit DNSs has important consequences for the understanding of $\gamma$-ray burst progenitors. We will try to estimate the birthrate of this new population in a forthcoming paper (Dewi, Pols \& van den Heuvel, in preparation).

\begin{figure}
 	 \centerline{\includegraphics[width=84mm]{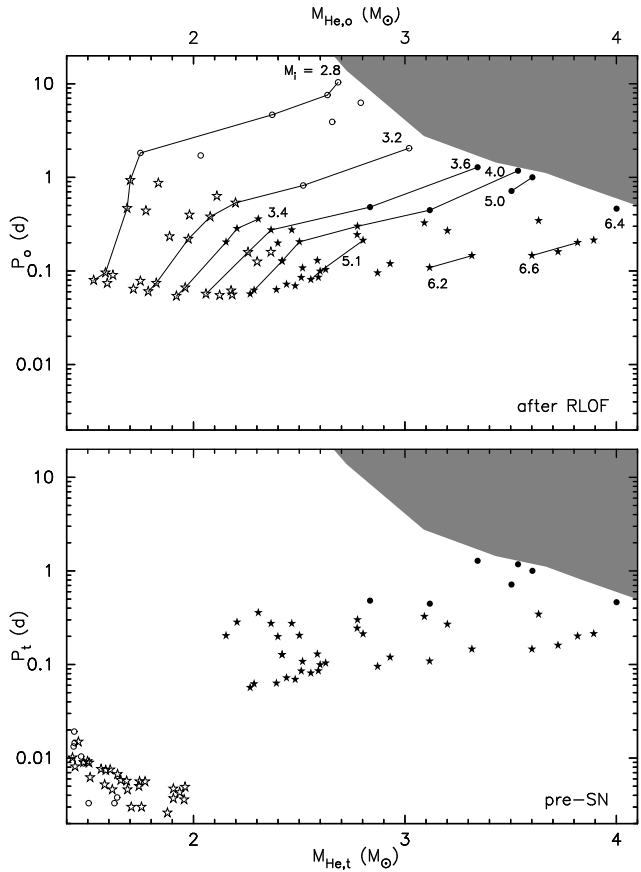}}
	 \caption[]{The masses and periods of the helium star-neutron star 
                    binaries at the end of our calculations. The upper panel 
                    gives the mass $M_{\mathrm{o}}$ and period $P_{\mathrm{o}}$ 
                    of the remnants before the spiral-in phase (after RLOF). The 
                    lower panel presents the mass $M_{\mathrm{t}}$ and period 
                    $P_{\mathrm{t}}$ after the spiral-in phase (prior to SN 
                    explosion). The star symbols represent the remnants of case BB 
                    mass transfer (see Paper~I), and circles indicate the remnant 
                    of case BC evolution from this work. The solid symbols represent 
                    the remnants of helium stars which do not go through a CE phase, and 
                    the open ones indicate those which do. Note that for systems 
                    that do not go through a CE phase (solid symbols), 
                    $P_{\mathrm{t}} = P_{\mathrm{o}}$ 
                    and $M_{\mathrm{t}} = M_{\mathrm{o}}$. In the upper panel, 
                    thin lines connect the remnants of helium stars with the same 
                    initial mass $M_{\mathrm{i}}$. The shaded area marks the region 
                    where a double neutron star can be produced by avoiding RLOF, 
                    taken from single helium stars calculation (Pols, in preparation) 
                    after taking into account the effect of stellar wind mass loss.}
	 \label{doublens:fig:remnant}
	\end{figure}

The upper panel of Fig.~\ref{doublens:fig:remnant} shows the masses and periods at the end of our calculations (i.e. post-RLOF). Systems which go through a subsequent CE phase are indicated by open symbols. If the neutron star in those systems completes the spiraling-in process, then their pre-SN (i.e. post-CE) masses and periods are presented in the lower panel. Since the explosion of the helium star can take place before the neutron star completes the spiral-in phase, the upper and lower panels of Fig.~\ref{doublens:fig:remnant} represent the maximum and minimum pre-SN mass and period, respectively, for systems which go through the CE phase. If the neutron star does not have enough time to complete the spiral-in before the helium star explodes, the pre-SN parameters are determined by the situation prior to the core collapse, i.e. mass and period will be between the values given in the upper and lower panels. However, Podsiadlowski (2001) showed that the most rapid orbital decay takes place at the beginning of the spiral-in phase. Hence, although the core collapse occurs before the spiral-in is terminated, the pre-SN period is likely to be close to the minimum. Note that for systems that do not go through a CE phase (solid symbols), the mass and period in the upper panel are the same as in the lower panel.

We have discussed in Sect.~\ref{doublens:subsec:collapse} the possibility of a dynamical ejection of the helium layer due to explosive neon flashes for $M_{\mathrm{He}} \sim 2.8 - 3.2$~\msun. Although it is unclear whether this actually occurs, and we ignore it in the remainder of the paper, we will discuss briefly how such a dynamical ejection might influence our results. If the ejection occurs before mass transfer becomes dynamically unstable, then we find the situation where the neutron star is orbiting the CO core, with mass as in the lower panel of Fig.~\ref{doublens:fig:remnant}, but with an orbit more like in the upper panel. If the ejection occurs after the spiral-in process is initiated, the future of the system depends on whether the ejection occurs after or before the completion of the spiral-in process. We will have the same situation as that where the core collapse occurs after the completion of the spiral-in (i.e. the lower panel) if the dynamical ejection of the helium layer takes place after the neutron star terminates the spiral-in phase. If the ejection ensues before the completion of the spiral-in, then again the CO core remains (with a mass as in the lower panel) but the period will be between the upper and lower panels.

\subsubsection{The type of supernova explosion}
\label{doublens:subsub:type}

The final amount of helium left in the envelope probably determines whether the explosion will be observed as a type Ib or a type Ic SN. The main observational criterion to distinguish between these types is the presence of helium in type Ib and its absence in type Ic SN. The conclusion drawn in Paper~I, that lower-mass helium stars and systems in close orbits are possible progenitors of type Ic SNs, and that higher-mass helium stars and systems in wide orbits produce type Ib SN, is still valid here.

We will discuss the possibility of the explosion type in the case of systems which undergo a CE and spiral-in phase. We consider the situation in which the SN explosion occurs after the neutron star completes the spiraling-in process, i.e. the whole helium envelope has been removed from the star. We discuss an extreme case, where the core collapses at the moment when the whole envelope is ejected. If the ejected matter is still surrounding the core, then the explosion will be observed as the helium-rich type Ib SN. This situation resembles the case where the explosion occurs inside the CE. 

In the other extreme, we have the situation where the explosion occurs after the end of the CE phase. In the case of the 2.8~\msun\ model with $P_{\mathrm{i}} = 2^{\mathrm{d}}$, if we assume that the CE phase lasts for only a few years (Sect.~\ref{doublens:subsubsec:decay}) and the explosion takes place at most a hundred years after the start of the CE phase (Sect.~\ref{doublens:subsubsec:explosion}), then the core collapses less than a hundred years after the CE phase is terminated. We assume that the envelope is ejected with the escape velocity, i.e. 420~\kms, and that during the SN explosion matter is ejected with a velocity of $\sim 10^{4}$~\kms. Using these velocities, we find that in 5 years the SN shell will catch up the envelope matter, forming a shock front. The SN shell might interact with the envelope matter in a form of a ring around the SN remnant like in SN~1987A, which was observed a year after the discovery of the SN. The explosion itself may be very dim since the ejected mass is very small (less than $\sim$ 0.5~\msun\ if we assume that the SN remnant has a mass of 1.4~\msun, see Table~\ref{doublens:tab:post-CE}). The models of 2.8~\msun\ with larger orbital periods have slightly higher masses but much larger radii and therefore lower escape velocities, such that the timescales for the SN shell to catch up the envelope matter are shorter ($\sim$ 2~yr in $P_{\mathrm{i}} = 10^{\mathrm{d}}$). In the models of 3.2~\msun, the escape velocities are large but the time left until the SN explosion is very short such that the interaction between the SN shell and the envelope matter takes place in $\sim$ 1.5~yr.


\section{The formation of double neutron star binaries}
\label{doublens:sec:doublens}

	\begin{figure}
 	 \centerline{\includegraphics[width=84mm]{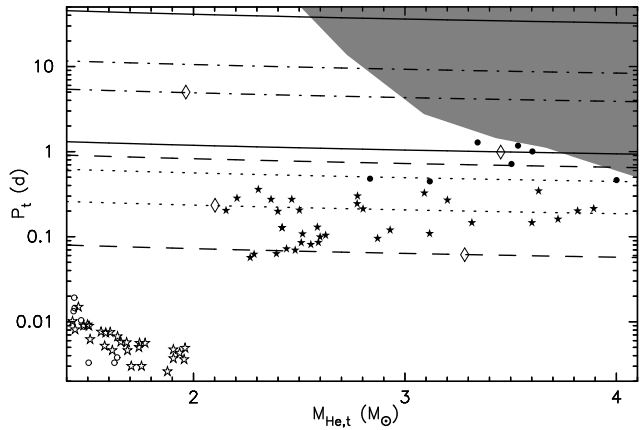}}
	 \caption[]{Range of allowed mass and period of the immediate pre-SN progenitor 
                    of double neutron-star binaries: B1913+16 (dashed-), B1534+12 
                    (dotted-), J1518+4904 (dash-dotted-), and J1811-1736 (solid-line). 
                    For each line style, upper and lower lines represent the maximum 
                    and minimum pre-SN periods, assuming that the ages of the pulsars 
                    are equal to their characteristic ages. The meaning of the symbols 
                    and shading are the same as in Fig.~\ref{doublens:fig:remnant}. 
                    Diamonds indicate the pre-SN masses and periods for the case of 
                    symmetric supernovae for all four pulsars.}
	 \label{doublens:fig:preSN}
	\end{figure}

	\begin{table*}  
         \caption[]{The orbital parameters of the observed galactic double 
                    neutron-star pulsars: the masses of the pulsar and its companion, 
                    the total mass of the system in \msun, orbital period in days, 
                    eccentricity, and characteristic age in $10^{8}$~yr.}
	 \label{doublens:tab:pulsar}
	 \begin{center}
	 \begin{tabular}{l@{\, \,}lll@{\, \,}r@{\, \,}ll@{\, \,}l}
	  \hline
	  \hline
          \noalign{\smallskip}
          PSR & ~~$M_{\mathrm{p}}$ & ~~$M_{\mathrm{c}}$ & ~~$M_{\mathrm{f}}$ & $P_{\mathrm{b}}$~~ & ~~~~$e$ & $\tau_{\mathrm{ch}}$ & Reference \\
          \noalign{\smallskip}
	  \hline
          \noalign{\smallskip}
          B1913+16   & 1.4411 & 1.3874 & 2.82843 &  0.323 & 0.61713 & 1.08 & Taylor \& Weisberg 1989,\\
                     &        &        &         &        &         &      & Taylor 1992\\
          B1534+12   & 1.3332 & 1.3452 & 2.67843 &  0.421 & 0.27368 & 2.4  & Wolszczan 1991,\\
                     &        &        &         &        &         &      & Stairs et al. 2002\\
          J1518+4904 & $1.56^{+0.13}_{-0.44}$ & $1.05^{+0.45}_{-0.11}$ & 2.62 &  8.634 & 0.24948 & $>160$ & Nice, Sayer \& Taylor 1996,\\
                     &                        &                        &      &        &         &        & Thorsett \& Chakrabarty 1999\\
          J1811-1736 & $< 2.3$ & $> 0.7$ & 2.6   & 18.779 & 0.828   & $9^{+4}_{-2}$ & Lyne et al. 2000 \\
          \noalign{\smallskip}
	  \hline
	  \hline
         \end{tabular}
	 \end{center}
	\end{table*}

Following up on our attempt in Paper~I, we will try to find constraints on the masses and separation of the progenitors of the observed galactic DNSs, and on the kick velocity that was imparted during the second SN explosion. This has been done before by various authors. However, as will be discussed later, here we reconsider their findings in the light of RLOF from helium stars which was not considered before. 

Table~\ref{doublens:tab:pulsar} lists the known galactic DNSs. Included in the table is the suggested DNS J1811-1736 (Lyne et al. 2000), although the possibility that the companion of the pulsar in this system is a main-sequence star or a red dwarf is still open (Mignani 2000). We plotted the allowed minimum and maximum pre-SN periods for each observed system in Fig.~\ref{doublens:fig:preSN}, which is expressed as (Flannery \& van den Heuvel 1975)
	\begin{eqnarray}
	(1-e_{\mathrm{f}})^{3} M_{\mathrm{f}} \, P_{\mathrm{f}}^{2} \leq 
	M_{\mathrm{s}} \, P_{\mathrm{t}}^{2} \leq 
	(1+e_{\mathrm{f}})^{3} M_{\mathrm{f}} \, P_{\mathrm{f}}^{2}
	\label{doublens:eq:post-per}	 
	\end{eqnarray}
Here $M_{\mathrm{s}}$, $M_{\mathrm{f}}$ are the total pre- and post-SN masses; $P_{\mathrm{t}}$, $P_{\mathrm{f}}$ are the pre- and post-SN periods. The values of $P_{\mathrm{f}}$ have been corrected for orbital decay during a time equal to the pulsar characteristic age (see Sect.~\ref{doublens:subsec:symmetry}). Eq.~(\ref{doublens:eq:post-per}) comes from the assumption that the radius of the pre-SN orbit (the helium star-neutron star binary) must be between the periastron and apastron distance of the post-SN orbit. 

\subsection{Previous studies}
\label{doublens:subsec:previous}

Yamaoka, Shigeyama \& Nomoto (1993) concluded that a symmetric explosion cannot explain the formation of B1534+12. Their conclusion was based on the argument that the pre-SN helium star mass is lower than 2.2~\msun, which is the critical mass for a helium star to become a neutron star (cf. Habets 1986). They argued that a kick velocity of 300 -- 460~\kms\ (for B1913+16) and 160 -- 260~\kms\ (for B1534+12) is required, assuming that the helium star does not fill its Roche lobe, which requires $M_{\mathrm{He}} > 5$~\msun\ in their adopted models . In the case where RLOF does occur they argue that the neutron star will spiral into the evelope of the helium star, such that the mass prior to SN explosion is the CO core mass which is smaller than the mass of the helium star. Since the required kick velocity increases with mass, a smaller kick velocity is needed for this case. Also in the situation where the helium star has experienced wind mass loss prior to the explosion, i.e. its final mass is much smaller than the initial helium star mass, a smaller kick velocity is required. 

Fryer \& Kalogera (1997) also found that a symmetric explosion in the formation of the observed DNSs requires a pre-SN separation smaller than the maximum radius of the helium star, i.e. RLOF must have occured. They assumed this must result in a CE phase. They found that, if a neutron star is able to accrete matter above its Eddington limit in the CE phase, the timescale needed for a neutron star to collapse into black hole is much smaller than the timescale for a helium star to evolve from its maximum radius up to the explosion. They concluded that the neutron star will collapse into a black hole. Hence, in this case, a symmetric explosion fails to explain the existence of the DNSs. Minimum kick velocities of 260, 220, and 50~\kms\ (for B1913+16, B1534+12, and J1518+4904, respectively) are required, from a progenitor with $a_{\mathrm{t}} \sim 4.5$~\rsun, $M_{\mathrm{t}} \sim 4.5$~\msun\ (for B1913+16 and B1534+12) and $a_{\mathrm{t}} \sim 30$~\rsun, $M_{\mathrm{t}} \sim 3$~\msun\ (for J1518+4904), in order to avoid RLOF from the helium star. As the completion to this work, Wex, Kalogera \& Kramer (2000) used the misalignment between the orbital angular momentum and the spin of B1913+16, together with its proper motion, to put more constraints on the kick velocity imparted during the second supernova.

None of the above-mentioned works was based on detailed calculations of helium stars in binary systems. Furthermore, most of them were carried out assuming that the helium star does not fill its Roche lobe. Based on our calculations we will reinvestigate the kick velocity imparted at the birth of the pulsar's companion, allowing for RLOF from the helium stars. With this assumption, the low pre-SN mass found by e.g. Yamaoka et al. (1993) is not necessarily the initial mass of the helium star. Hence, we allow lower pre-SN masses. Also by allowing the helium star to fill its Roche lobe without causing the neutron star to collapse into a black hole, progenitors in closer orbits than those found by Fryer \& Kalogera (1997) are possible.

In Fig.~\ref{doublens:fig:preSN} we compare the allowed range of pre-SN orbital periods of the DNSs with the results of our model calculations, reproduced from Fig.~\ref{doublens:fig:remnant} (lower panel). Fig.~\ref{doublens:fig:preSN} shows that the possible progenitors of the short-orbit DNSs B1913+16 and B1534+12 are helium stars with $M_{\mathrm{t}} > 2.2$~\msun, i.e. with initial mass $M_{\mathrm{He}} > 3.3$~\msun\ (solid symbols), if indeed helium stars of lower mass leave DNSs with very small periods (open symbols). This means their progenitors were main-sequence stars more massive than 12~\msun\ which underwent case B evolution, or stars more massive than 10~\msun\ which experienced case C mass transfer.

\subsection{Symmetric explosion}
\label{doublens:subsec:symmetry}

	\begin{figure}
 	 \centerline{\includegraphics[width=84mm]{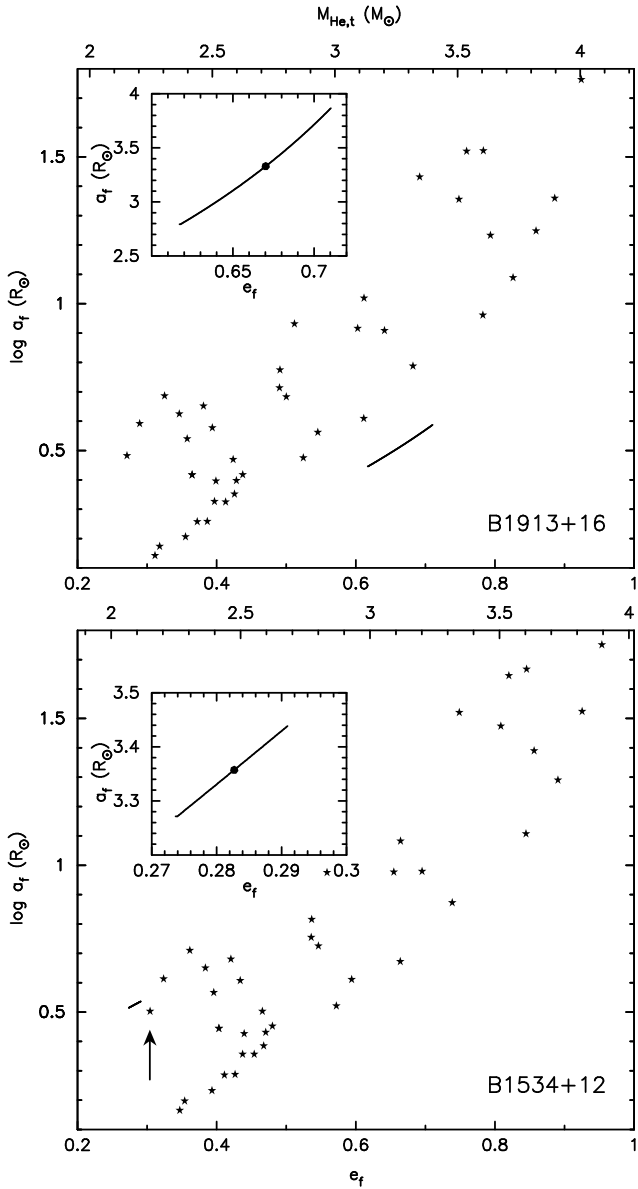}}
	 \caption[]{The evolution of eccentricity and separation of B1913+16 and B1534+12
                    (solid lines) back from the present time to twice their characteristic
                    ages (the values at the age equal to the characteristic age are indicated
                    by the solid circle in the inset of each panel). Also plotted here, in 
                    solid stars, are the post-SN eccentricity and separation resulting from 
                    our calculations, assuming a symmetric supernova explosion. The pre-SN 
                    mass is indicated at the top x-axis.}
	 \label{doublens:fig:symmetry}
	\end{figure}

In this section we reinvestigate whether a symmetric explosion can explain the observed DNSs, if the helium stars are allowed to undergo RLOF. By means of eqs.~(\ref{doublens:eq:gwr-sep}) and (\ref{doublens:eq:gwr-ecc}) we calculate the evolution of eccentricity and separation of the DNSs B1913+16 and B1534+12 due to gravitational-wave radiation, back from the present time to twice their characteristic ages, as presented in Fig.~\ref{doublens:fig:symmetry} as solid-line segments. The solid circle represents the eccentricity and separation for an age equal to the characteristic age. In a symmetric SN explosion, the post-SN eccentricity, $e_{\mathrm{f}}$, and separation, $a_{\mathrm{f}}$, are related to the pre-SN mass, $M_{\mathrm{t}}$, and separation, $a_{\mathrm{t}}$, as (Hills 1983)
	\begin{eqnarray}
         M_{\mathrm{t}} = e_{\mathrm{f}} \, (M_{\mathrm{p}} + M_{\mathrm{c}}) + M_{\mathrm{c}}
         \label{doublens:eq:pre-ecc}	 
	\end{eqnarray}
	\begin{eqnarray}
         \frac{a_{\mathrm{t}}}{a_{\mathrm{f}}} = 
         \frac{2 \, M_{\mathrm{c}} + M_{\mathrm{p}} - M_{\mathrm{t}}}
              {M_{\mathrm{p}} + M_{\mathrm{c}}}
         \label{doublens:eq:pre-sep}	 
	\end{eqnarray}
where $M_{\mathrm{p}}$ and $M_{\mathrm{c}}$ are the masses of the pulsar and its companion, respectively. The star symbols in Fig.~\ref{doublens:fig:symmetry} represent the expected $e_{\mathrm{f}}$ and $a_{\mathrm{f}}$ for each of our model calculations, calculated from eqs.~(\ref{doublens:eq:pre-ecc}) and (\ref{doublens:eq:pre-sep}).

The upper panel of Fig.~\ref{doublens:fig:symmetry} shows that in the case of a symmetric explosion, B1913+16 should have formed from a helium star with $M_{\mathrm{t}} = 3.13 - 3.42$~\msun, depending on its true age (which is assumed to be less than twice its characteristic age). However, the separation is smaller than the minimum separation allowed for dynamically stable RLOF (cf. fig.~10 of Paper~I). This implies that B1913+16 cannot be formed by assuming a symmetric explosion from helium star-neutron star binaries which undergo a mass-transfer phase, regardless of the exact age of the system.

The eccentricity of B1534+12 corresponds to $M_{\mathrm{t}} = 2.07 - 2.12$~\msun. The eccentricity and separation of this pulsar lie close to the remnants of our 3.4~\msun\ model (indicated by an arrow in the lower panel of Fig.~\ref{doublens:fig:symmetry}). However, we have discussed in Sect.~\ref{doublens:subsec:caseBB} that the future of this particular mass is not clear. These systems may also undergo a CE and spiral-in phase and move to the lower-left part of the plane as in the case of lower-mass helium stars; and in that case cannot be considered as a possible progenitor of B1534+12. Hence, a conclusion that a symmetric explosion can explain the formation of B1534+12 is marginal. Our results confirm previous work that it is more likely that both B1913+16 and B1534+12 are formed by an asymmetric SN explosion, although the constraints on the kick velocity become much weaker if we allow for RLOF from the helium star -- as will be discussed in Sect.~\ref{doublens:subsec:asymmetry}. 

The pre-SN mass and period ($M_{\mathrm{t}}$, $P_{\mathrm{t}}$) required in a symmetric explosion are indicated by diamond symbols in Fig.~\ref{doublens:fig:preSN}, for each of the four DNSs, assuming an age equal to the characteristic age. We find $M_{\mathrm{t}} =$ 1.96~\msun\ as the pre-SN mass of the wide DNS J1518+4904 if we assume a symmetric explosion. This mass (with $a_{\mathrm{t}} = a_{\mathrm{f}} (1-e_{\mathrm{f}})$) is located in the region where helium stars of $\sim$ 2.8 -- 2.9~\msun\ will undergo a CE and spiral-in phase (compare Figs.~\ref{doublens:fig:remnant} and \ref{doublens:fig:preSN}). Assuming that they complete the spiraling-in process, their position in the pre-SN mass-period plane will move to the lower-left part of Fig.~\ref{doublens:fig:preSN}. Even if they undergo a SN before completing the spiral-in, their final orbits are likely to be much closer than those at the onset of the CE phase. We conclude that a symmetric explosion cannot explain the formation of J1518+4904 and that it can only be produced from a helium star more massive than 2.8~\msun\ which avoids RLOF (shaded area in Fig.~\ref{doublens:fig:preSN}), i.e. main-sequence stars more massive than 10~\msun\ in relatively wide orbits. 

Accurate masses of the components of J1811-1736 are not known yet. For our calculations we assume that the components have equal masses. With a symmetric explosion, we find $M_{\mathrm{t}} = 3.45$~\msun, implying that the progenitor of J1811-1736 is a helium star of about 4.0~\msun. Helium stars with initial masses $\sim$ 3.3 -- 4.0~\msun\ in wide orbits which undergo marginal RLOF can be the progenitors of J1811-1736. However, because of the much larger parameter space, a progenitor that avoids RLOF is more likely. Whether a symmetric explosion can explain the formation of J1811-1736 depends on the exact masses of the components which in turn determine the pre-SN $a_{\mathrm{t}}$ and $M_{\mathrm{t}}$. 

\subsection{Asymmetric explosion}
\label{doublens:subsec:asymmetry}

	\begin{figure}
 	 \centerline{\includegraphics[width=84mm]{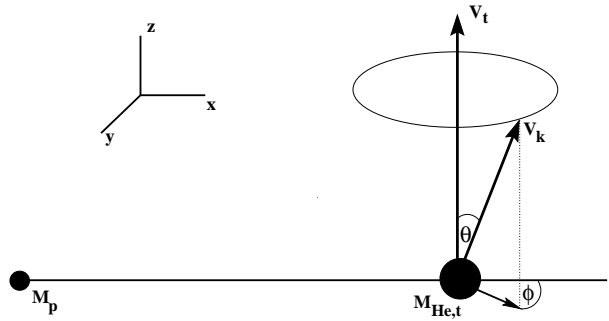}}
	 \caption[]{The orientation of the kick velocity $V_{\mathrm{k}}$
                    relative to the original orbital velocity
                    $V_{\mathrm{t}}$.}
	 \label{doublens:fig:vkick}
	\end{figure}

If the kick velocity $V_{\mathrm{k}}$ makes an angle $\theta$ with respect to the pre-explosion orbital velocity $V_{\mathrm{t}}$, then we can write the relation between the pre-SN and post-SN separations, $a_{\mathrm{t}}$ and $a_{\mathrm{f}}$, as (Hills 1983)
	\begin{eqnarray}
         \frac{a_{\mathrm{t}}}{a_{\mathrm{f}}} = 2 -
         \frac{M_{\mathrm{s}}}{M_{\mathrm{f}}}
         \left[ 1 + \nu^{2}+ 2 \nu \cos \theta \right]
         \label{doublens:eq:kicksep}	 
	\end{eqnarray}
where $M_{\mathrm{s}} = M_{\mathrm{t}} + M_{\mathrm{p}}$, $M_{\mathrm{f}} = M_{\mathrm{c}} + M_{\mathrm{p}}$, $\nu = V_{\mathrm{k}}/V_{\mathrm{t}}$ and $V_{\mathrm{t}}^{2} = G \, M_{\mathrm{s}} / a_{\mathrm{t}}$.  If we rotate $V_{\mathrm{k}}$ around $V_{\mathrm{t}}$ in a cone of apex angle $\theta$ as is shown in Fig.~\ref{doublens:fig:vkick}, then $\phi$ is the location of $V_{\mathrm{k}}$ on this cone, such that $\phi = 0$ corresponds to $V_{\mathrm{k}}$ in the original orbital plane and has a Cartesian component pointing radially outward from the focus of the orbit (Hills 1983). The eccentricity of the post-SN orbit is given by
	\begin{eqnarray*}
         G \, a_{\mathrm{f}} \, M_{\mathrm{f}} \, (1 - e_{\mathrm{f}}^{2}) =
         a_{\mathrm{t}}^{2} [V_{\mathrm{k}}^2 \, \sin^{2}\!\theta \sin^{2}\!\phi +
                             (V_{\mathrm{k}} \, \cos \theta + V_{\mathrm{t}})^{2}]
        \end{eqnarray*}
or
	\begin{eqnarray}
         1 - e_{\mathrm{f}}^{2} = 
         \frac{a_{\mathrm{t}}}{a_{\mathrm{f}}} \, 
         \frac{M_{\mathrm{s}}}{M_{\mathrm{f}}} \,
         [ 1 + 2 \nu \cos \theta + 
         \nu^{2} (\cos^{2}\!\theta + \sin^{2}\!\theta \sin^{2}\!\phi) ]
         \label{doublens:eq:kickecc}	 
	\end{eqnarray}
After the explosion, a binary will remain bound if the right-hand side of eq.~(\ref{doublens:eq:kicksep}) is positive. Hence, after an asymmetric explosion with a kick velocity $V_{\mathrm{k}}$, a binary with pre-SN parameters ($M_{\mathrm{s}}$, $a_{\mathrm{t}}$) will not be disrupted if the angle $\theta$ is larger than a critical angle, i.e.
	\begin{eqnarray}
         \theta > \theta_{\mathrm{cr}} =
         \cos^{-1} \left[ \frac{2 M_{\mathrm{f}} - M_{\mathrm{s}}(\nu^{2}+1)}
                               {2 \nu M_{\mathrm{s}}} \right]
         \label{doublens:eq:theta}	 
	\end{eqnarray}
Since $\cos \theta_{\mathrm{cr}} \geq -1$ an absolute maximum kick velocity can be derived from the above equation as
	\begin{eqnarray}
         V_{\mathrm{k,max}} = (1 + \sqrt{2 \, M_{\mathrm{f}}/M_{\mathrm{s}}}) \,\, V_{\mathrm{t}}
         \label{doublens:eq:vmax}	 
	\end{eqnarray}
A kick with this magnitude has to be directed opposite to the orbital motion ($\theta = 180\degr$). For $M_{\mathrm{s}} > 2 \, M_{\mathrm{f}}$ there is a minimum kick velocity 
	\begin{eqnarray}
         V_{\mathrm{k,min}} = (1 - \sqrt{2 \, M_{\mathrm{f}}/M_{\mathrm{s}}}) \,\, V_{\mathrm{t}}
         \label{doublens:eq:vmin}	 
	\end{eqnarray}
which also has to be directed at $\theta = 180\degr$. A $V_{\mathrm{k,min}} = 0$ requires $M_{\mathrm{s}} \leq 2 \, M_{\mathrm{f}}$ as derived by Blaauw (1961) for a symmetric explosion.

For a given observed DNS, we can find the pre-SN parameters by means of eqs.~(\ref{doublens:eq:kicksep}) and (\ref{doublens:eq:kickecc}) for a certain kick velocity ($V_{\mathrm{k}}, \theta, \phi$). The kick direction is constrained by $0 \leq \sin^{2}\!\phi \leq 1$. Independent of the magnitude of the kick velocity, the limit $\sin^{2}\!\phi = 1$ gives eq.~(\ref{doublens:eq:post-per}) which can also be written as  
	\begin{eqnarray}
         a_{\mathrm{f}} \, (1-e_{\mathrm{f}}) \leq a_{\mathrm{t}} 
         \leq a_{\mathrm{f}} \, (1+e_{\mathrm{f}})
         \label{doublens:eq:sinphi}	 
	\end{eqnarray}
These lower and upper limits in separation are presented as the thick horizontal lines in Fig.~\ref{doublens:fig:kick}, for the two closest DNSs. The limit $\sin^{2}\!\phi = 0$ for different kick velocity magnitudes is given by the thin lines. For a given kick velocity, the possible pre-SN parameters lie between the lines of $\sin^{2}\!\phi = 1$ and that of $\sin^{2}\!\phi = 0$ (i.e. inside the shaded area marked by a certain kick magnitude).

\subsubsection{Formation of DNS without a mass transfer phase}
\label{doublens:subsubsec:noRLOF}

\begin{figure}
 	 \centerline{\includegraphics[width=84mm]{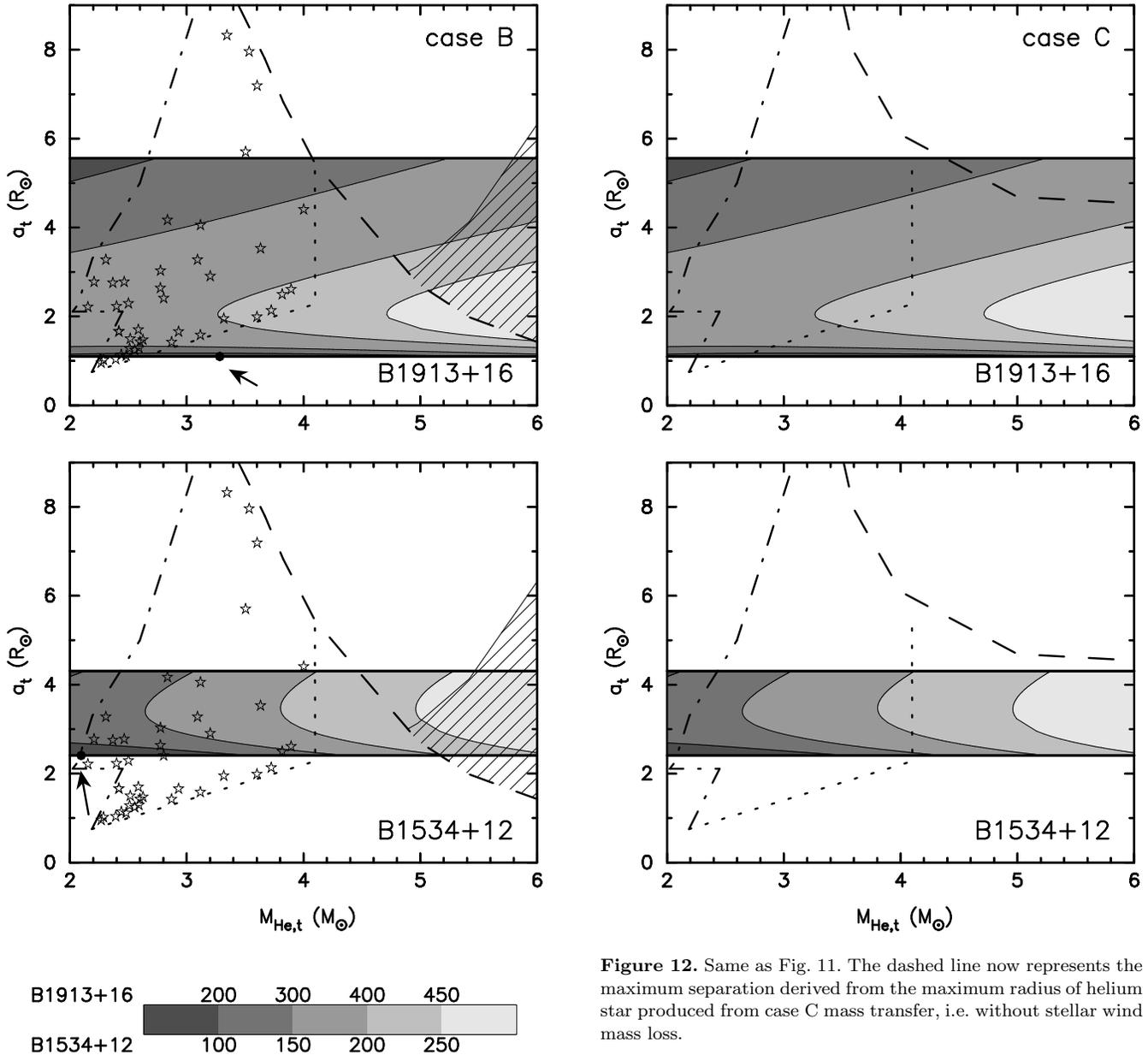}}
	 \caption[]{The possible magnitude of the kick velocity imparted at
                    the birth of the second neutron star in B1913+16 (upper) and B1534+12
                    (lower panel), with the assumption that the pulsar age is the same as
                    its characteristic age. The star symbols represent the pre-SN 
                    parameters of our case BB and BC calculations which do not go through 
                    a CE phase. Thick horizontal lines mark the minimum and maximum pre-SN 
                    separations constrained by $\sin^{2}\!\phi = 1$. Thin lines are 
                    obtained from the constraint of $\sin^{2}\!\phi = 0$. The solid circle 
                    (indicated by an arrow) gives the pre-SN parameter if the explosion is 
                    symmetric ($V_{\mathrm{k}} = 0$). Below and to the right of the dotted 
                    line, mass transfer 
                    occurs on the dynamical timescale leading to a merger. To the left of the 
                    dash-dotted line is the region where mass transfer ends in a CE phase. 
                    The dashed line represents the maximum separation derived from the 
                    maximum radius of a helium star produced from case B mass transfer, after 
                    taking into account the stellar wind mass loss. The hatched area 
                    indicates the region where the separation is smaller than the minimum
                    separation derived from the helium zero-age main sequence. The 
                    corresponding kick velocities for each panels are presented at the 
                    bottom of the figure.}
	 \label{doublens:fig:kick}
	\end{figure}

We will first revisit the investigation of the formation of the DNSs assuming that the progenitors do not experience RLOF. A helium star will not fill its Roche lobe if the Roche radius is larger than its maximum radius, which defines a critical orbital separation $a_{\mathrm{max}}$. We plot this separation as the dashed line in Fig.~\ref{doublens:fig:kick}. This maximum separation is derived from the maximum radius taking into account wind mass loss as defined in eq.~(\ref{doublens:eq:wind}). This represents the situation where the helium star is formed after case B mass transfer, and has been losing mass by stellar winds during its evolution prior to RLOF. A minimum separation, below which a helium star-neutron star binary cannot be formed, is defined by equating the Roche radius to the helium zero-age main sequence radius. Due to the wind mass loss, this minimum separation increases according to eq.~(\ref{doublens:eq:orbit}) with $\alpha = 1$ and $\beta = 0$. We plot the region with separation less than the minimum separation at the end of the evolution of a helium star against its final mass as the hatched area. Helium star-neutron star binaries cannot be formed in this part of the diagram. These constraints leave the region above the maximum (dashed line) and minimum (hatched area) separations as the allowed pre-SN parameters in Fig.~\ref{doublens:fig:kick} (however, these constraints depend very much on the choice of the wind mass-loss rate).

Without RLOF from the helium stars, we find the same minimum kick velocity as Fryer \& Kalogera (1997), i.e. 260 and 220~\kms\ for B1913+16 and B1534+12, respectively (see the shaded area above the dashed line in Fig.~\ref{doublens:fig:kick}). The kick must be directed backwards, with $\theta > 130\degr$. For a low $V_{\mathrm{k}}$ ($\la$ 425~\kms\ for B1913+16 and $\la$ 275~\kms\ for B1534+12) a kick in any azimuthal direction $\phi$ is allowed to form the binaries with the observed parameters, but the higher $V_{\mathrm{k}}$ the more restricted the allowed orientation of the kick.

The inclusion of the minimum separation (hatched area in Fig.~\ref{doublens:fig:kick}) as an additional constraint, which has not been taken into account in previous works, enables us to determine the maximum pre-SN mass. The maximum pre-SN separation of B1913+16 intersects the line of $a_{\mathrm{t}} = a_{\mathrm{min}}$ at $M_{\mathrm{t}} \sim$ 5.8~\msun, which is the final mass of a $\sim$ 16~\msun\ helium star after stellar wind mass loss. This means that B1913+16 cannot have formed from a helium star initially more massive than 16~\msun. Similarly, we derive $\sim$ 13~\msun\ as the upper limit for the helium star progenitor of B1534+12. The maximum pre-SN mass also implies a maximum kick velocity, from eq.~(\ref{doublens:eq:vmax}), which is 1\,230~\kms\ for both systems.

In discussing the above situation, we consider that the helium star-neutron star binaries are produced at the time when the helium star is on its zero-age main sequence, i.e. the remnant of a Be/X-ray binary which went through a CE phase initiated in case B. If the CE phase is initiated in case C (i.e. after the termination of helium core burning), the helium core does not experience significant wind mass loss and therefore has a maximum radius equal to that of a helium star evolving without wind mass loss. This maximum radius is plotted as the dashed line in Fig.~\ref{doublens:fig:caseC}. For case C remnants there is no minimum separation which is constrained by the helium zero-age main sequence radius. Whether RLOF can occur or not depends on the situation at the cessation of the CE phase which is very uncertain. We will consider the case where the case C remnant does not fill its Roche lobe (the area above the dashed line in Fig.~\ref{doublens:fig:caseC}). This area is located outside the range of possible pre-SN periods of B1534+12. Therefore we conclude that B1534+12 cannot be formed by case C mass transfer from Be/X-ray binaries without a further RLOF phase. To produce B1913+16 from this type of remnant, a very high $\theta > 150\degr$ is needed. The upper limit to the pre-SN mass, and the maximum kick velocity, are restricted by the threshold mass for black hole formation, which is still uncertain.

For the two wide-orbit DNSs, J1518+4904 and J1811-1736, the maximum separation for RLOF for case C remnants coincides with case B, because helium stars of such mass ($2.5 \la M_{\mathrm{t}}/{\mathrm{M_{\sun}}} \la 4.0$, see Fig.~\ref{doublens:fig:remnant}) lose very little mass in a wind. Hence, the minimum kick velocities are the same in both cases, i.e. 50~\kms\ for J1518+4904, as also found by Fryer \& Kalogera (1997), and 10~\kms\ for J1811-1736 (the latter depending on the uncertain component masses for this system). The maximum kick velocity, again, depends on the threshold mass for black hole formation.

\subsubsection{Formation of DNS with RLOF}
\label{doublens:subsubsec:RLOF}

	\begin{figure}
 	 \centerline{\includegraphics[width=84mm]{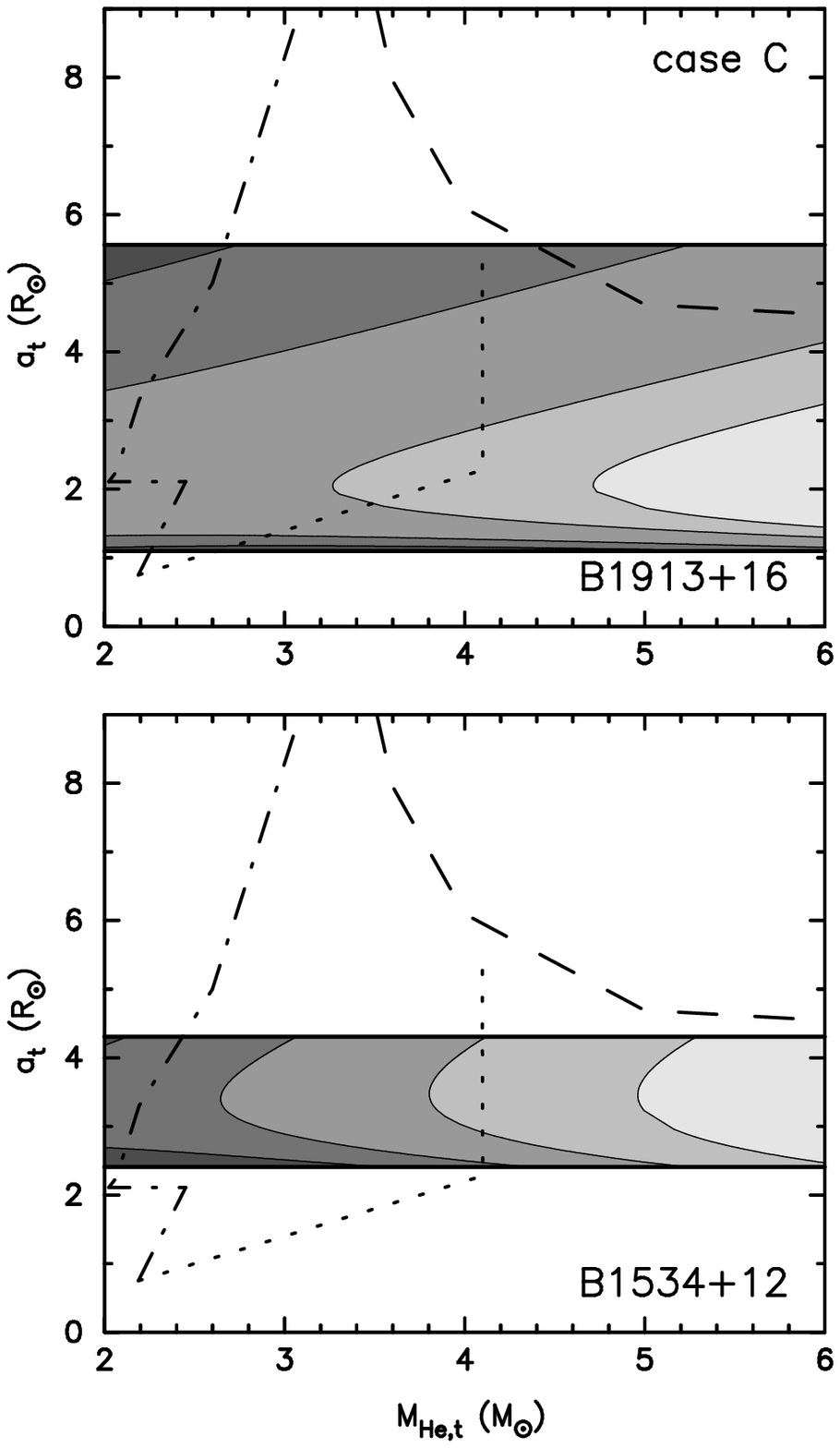}}
	 \caption[]{Same as Fig.~\ref{doublens:fig:kick}. 
                    The dashed line now represents the maximum separation derived from the 
                    maximum radius of helium star produced from case C mass transfer, i.e.
                    without stellar wind mass loss.}
	 \label{doublens:fig:caseC}
	\end{figure}

The region in the ($M_{\mathrm{t}}$, $a_{\mathrm{t}}$) plane where close-orbit DNSs can be formed through RLOF is limited, as shown in Fig.~\ref{doublens:fig:kick}. It is bounded at small separations because mass transfer becomes dynamically unstable leading to a merger (below the dotted line). In Paper~I we found that mass transfer from $M_{\mathrm{He}} > 6.7$~\msun\ is dynamically unstable leading to a CE phase. This range of mass is shown as the region to the right of the dotted line (with $M_{\mathrm{t}} \ga 4$~\msun). We have not carried out calculations for this range of mass. However, we suggest that such a CE phase would lead to a merger or a formation of DNS in tight orbit, depending on the binding energy of the envelope. At low $M_{\mathrm{t}}$, to the left of the dash-dotted line, systems experience a CE phase after RLOF. This leaves the region between the dotted, dashed, and dash-dotted line, in which the results of our RLOF calculations lie (the star symbols). We do not present the star symbols in Fig.~\ref{doublens:fig:caseC} because the calculations we have done are the remnants of case B mass transfer. Although we have not computed the evolution of the remnants of case C mass transfer, we expect that the limits for dynamically unstable mass transfer (dotted line) and for systems undergoing a CE phase after RLOF (dashed-dotted line) are the same as for the remnants of case B evolution.

We find that B1913+16 can be formed by an asymmetric explosion with minimum kick velocity of 70~\kms, which must have been directed along the orbital motion ($\theta < 20\degr$). A kick velocity as low as 10~\kms\ is enough to produce B1534+12, which requires $\theta > 85\degr$. A large $V_{\mathrm{k}}$ requires $\theta$ close to $180\degr$, and can only be imparted in a very close orbit ($a_{\mathrm{t}} \sim a_{\mathrm{f}} \, (1-e_{\mathrm{f}})$, cf. eq.~(\ref{doublens:eq:kicksep})). With this $a_{\mathrm{t}}$ and using eq.~(\ref{doublens:eq:vmax}) we find 1\,810~\kms\ and 1\,310~\kms\ as the maximum kick velocities that can produce B1913+16 and B1534+12, respectively. To produce a system with post-SN parameters ($M_{\mathrm{f}}$, $a_{\mathrm{f}}$) from a binary with pre-SN parameters ($M_{\mathrm{s}}$, $a_{\mathrm{t}}$), it can be seen from eq.~(\ref{doublens:eq:kickecc}) that with the same kick velocity, $e_{\mathrm{f}}$ decreases with $\theta$. Since B1534+12 has almost the same mass and separation as B1913+16 but lower eccentricity, in general, with the same kick velocity, a larger angle $\theta$ is needed to produce B1534+12 than that to form B1913+16.

We have discussed in Sect.~\ref{doublens:subsubsec:explosion} that there are two open possibilities for the fate of helium stars which go through a CE phase. If the core collapses after the spiral-in phase is completed, then the system becomes a double neutron star in a tight orbit, i.e. the situation is as described in the previous paragraph. If the helium star explodes while the neutron star is still spiraling-in, the pre-SN mass and separation will be somewhere between $M_{\mathrm{o}}$ and $M_{\mathrm{t}}$ and between $P_{\mathrm{o}}$ and $P_{\mathrm{t}}$ (cf. the upper and lower panels of Fig.~\ref{doublens:fig:remnant}), i.e. the region to the left of the dash-dotted line in Fig.~\ref{doublens:fig:kick} may not be empty. The inclusion of this area with lower mass would allow a wider range of kick velocities and directions. The magnitude and orientation of the minimum kick velocity for B1913+16 remains the same. In this situation B1534+12 can be produced by a symmetric explosion or with very low kick velocity in all directions $\theta$ and $\phi$.

Comparing our results with those of Yamaoka et al. (1993), Fryer \& Kalogera (1997), Wex et al. (2000), and Sect.~\ref{doublens:subsubsec:noRLOF}, we can see that by allowing stable RLOF from the helium star, systems with lower pre-SN mass and lower pre-SN separation than derived in previous works can also be possible progenitors of B1913+16 and B1534+12. We have shown that a low kick velocity can then explain the formation of these DNSs.


\section{Conclusions}
\label{doublens:sec:conclusions}

We have evolved helium stars with masses in the interval 2.8 -- 6.4~\msun\ with a 1.4~\msun\ neutron-star companion, in which the helium star fills its Roche lobe during carbon core burning or beyond (case BC mass transfer). This is the completion to our earlier work, i.e. mass transfer during helium core burning (case BA) and helium shell burning (case BB evolution). We studied the late stage of evolution of the helium star-neutron star binaries as well as the possible remnants of the systems.

Case BB and BC mass transfer in helium stars of 2.8 -- 3.3~\msun\ as well as from $3.3 < M_{\mathrm{He}}/{\mathrm{M_{\sun}}} < 3.8$ in very close orbits ($P \la 0\fd25$), will end up in a CE phase towards the end of their evolution, just before the expected SN explosion. These systems originate from main-sequence stars with masses of 10 -- 12~\msun\ which underwent case B evolution, or 9 -- 10~\msun\ which experienced case C mass transfer. We found that all systems are able to survive the spiral-in phase, producing very tight DNSs with $P \sim 0\fd01$. These systems probably will not be observed due to their very short merger timescale, and would have important consequences for the detection rate of gravitational-wave radiation and for the understanding of $\gamma$-ray burst progenitors. On the other hand, there is a possibility that the helium star will explode before the neutron star completes the spiraling-in process, resulting in a SN explosion inside a CE. 

For $M_{\mathrm{He}} > 3.3$~\msun\ or $M_{\mathrm{He}} > 3.8$~\msun\ in close orbits, we conclude that a CE and spiral-in phase does not occur. These systems will produce DNSs with periods of $0\fd1 - 1^{\mathrm{d}}$, which suggests they are candidate progenitors of B1913+16 and B1534+12. The pre-SN mass is larger than 2.2~\msun. These systems originate from main-sequence stars more massive than 12~\msun\ which underwent case B evolution, or more massive than 10~\msun\ which experienced case C mass transfer. 

We have also studied the second SN explosion to investigate whether a kick velocity is required at the birth of the young neutron star. DNS B1913+16 cannot be formed by a symmetric explosion. A minimum kick of 70~\kms\ in the direction of the orbital velocity is needed to produce the observed orbital parameters. A symmetric explosion can explain the formation of B1534+12 only if the progenitor, a 3.4~\msun\ helium star, avoids a CE and spiral-in phase. Our calculations are inconclusive on this issue. A higher mass progenitor needs a minimum kick velocity of 10~\kms\ in the direction opposite to the orbital velocity.

The wide-orbit DNS J1518+4904 (and probably J1811-1736) can only be produced from helium star-neutron star systems which did not go through a mass-transfer phase; with the helium stars more massive than 2.5~\msun, i.e. main-sequence stars more massive than 10~\msun\ in relatively wide orbit. Without a mass-transfer phase, an asymmetric explosion with a minimum kick velocity of 50 and 10~\kms\ (for J1518+4904 and J1811-1736) is needed.

\section*{Acknowledgements}
This work was sponsored by NWO Spinoza Grant 08-0 to E.~P.~J. van den Heuvel. It is a pleasure to thank Ed van den Heuvel and Norbert Langer for various discussions; Stan Woosley and Alexander Heger for invaluable help concerning the late stage of stellar evolution; Natasha Ivanova and Krzysztof Belczy{\' n}ski for their critical comments on the manuscript. Comments from anonymous referee help in improving the presentation of the manuscript.

\label{lastpage}

\end{document}